\documentstyle[11pt,rtrpp4]{article}
\def\epsfsize#1#2{\hsize}

\def\alwaysmath#1{\ifmmode{#1}\else{$#1$}\fi}
\def\msun{\alwaysmath{\,{\cal M}_{\odot}}}
\def\lsun{\alwaysmath{\,{\cal L}_{\odot}}}

\def\logteff{\alwaysmath{\log\,T_{\rm eff}}}
\def\etal{{et al.~}}

\def\feh{{\rm [Fe/H]}}
\def\mh{{\rm [M/H]}}
\def\alphafe{{\rm [\alwaysmath{\alpha}/Fe]}}

\def\fei{{\rm Fe}\,{\sc i}}
\def\feii{{\rm Fe}\,{\sc ii}}

\def\etal{{et al.~}}

\def\C12C13{$^{12}$C/$^{13}$C}

\def\d14{$\Delta V_{1.4}$ }

\slugcomment{ In press in the AJ}

\lefthead{F.R.Ferraro \etal}
\righthead{The RGB, AGB and HB of GGCs}
\begin{document}

 \title
 {THE GIANT, HORIZONTAL AND ASYMPTOTIC BRANCHES OF GALACTIC GLOBULAR
CLUSTERS. I.
THE CATALOG, PHOTOMETRIC OBSERVABLES AND FEATURES}

\author
{
F.R.~Ferraro\altaffilmark{1,2}, 
M. Messineo\altaffilmark{2},   
F. Fusi~Pecci\altaffilmark{2,3},  
M.A. De Palo\altaffilmark{4},}   

\author{O. Straniero\altaffilmark{5},
A. Chieffi\altaffilmark{6},
M. Limongi\altaffilmark{7}}

\altaffiltext{1}{European Southern Observatory, Karl Schwarzschild 
Strasse, 2, D-85748 Garching bei Munchen, GERMANY}
\altaffiltext{2}{on leave from Osservatorio Astronomico di Bologna, via Zamboni 33, I-40126
Bologna, ITALY}
\altaffiltext{3}{Stazione Astronomica di Cagliari, 09012 Capoterra, ITALY }
\altaffiltext{4}{Dipartimento di Astronomia, Universit\`a  di Bologna, I-40126
Bologna, ITALY }
\altaffiltext{5}{Osservatorio Astronomico di Collurania, via M. Vaggini,  I-64100
Teramo, ITALY }
\altaffiltext{6}{Istituto di Astrofisica Spaziale del CNR, CP 67, I-00044
Frascati, ITALY }
\altaffiltext{7}{Osservatorio Astronomico di Monte Porzio,
I-00040 Monte Porzio Catone, Italy}

\begin{abstract}
A catalog including a set of 
the most recent Color Magnitude Diagrams (CMDs)
is presented for a sample of 61 Galactic Globular Clusters (GGCs).
We used this data-base to perform an homogeneous systematic
analysis of the evolved sequences (namely,
Red Giant Branch (RGB), Horizontal Branch (HB) and Asymptotic
Giant Branch (AGB)).
Based on this analysis, we present:
(1) a new procedure to measure the level of the $ZAHB$ ($V_{ZAHB}$)
 and an homogeneous set of distance moduli obtained adopting the
HB as {\it standard candle};
(2) an independent estimate for RGB metallicity indicators
and  new calibrations of these parameters
in terms of both spectroscopic ($\feh_{\rm CG97}$) and
global metallicity ($\mh$, including 
also the $\alpha-$elements enhancement).
The set of equations presented can be used to simultaneously derive
a {\it photometric} estimate of the metal abundance and the reddening
from the morphology and the location of the RGB in the $(V,B-V)$-CMD.
(3) the location of the RGB-Bump (in 47 GGCs) and the AGB-Bump
(in 9 GGCs). The dependence of these features on the metallicity
is discussed. We find that by using the latest theoretical models
and the new metallicity scales the earlier discrepancy between 
theory and  observations ($\sim 0.4$ mag) completely disappears.
\end{abstract}


\section{Introduction}
Stellar evolutionary models are often used to infer relevant properties of the
Galaxy and the early Universe; for this reason the check of their adequacy and
accuracy can be regarded as a {\it pivotal} project of the modern astrophysical 
research (Renzini \& Fusi Pecci 1988).

The advent of the charge-coupled device (CCD) and, more recently,  the 
availability of the {\it Hubble Space Telescope}, supported by
the modern highly powerful software for photometric data analysis in crowded
fields, have greatly enhanced the possibility of using the Galactic Globular 
Clusters (GGCs) as the ideal laboratory to test the stellar evolution theories.

Within this framework, we started a long term project devoted to carry out 
a detailed quantitative analysis of the evolved sequences
(namely, Red Giant Branch, Horizontal Branch, Asymptotic Giant Branch, 
hereafter RGB, HB, AGB, respectively) in the Colour Magnitude Diagram 
(CMD) of GGCs.

The methodological approach of our study has been presented in a
series of papers concerning the photometry of wide samples 
of stars in a selected set of GGCs (see for example Ferraro et al. 1990, 
1991, 1992a,b, 1993, 1994, 1995, 1997a and Buonanno et al. 1994).
Some results on  specific sequences  can be found in  Fusi Pecci et al.
1990 (hereafter F90) and Ferraro 1992 (for the RGB), 
Fusi Pecci et al. 1992, 1993, Buonanno et al 1997, 
Ferraro 1997b, 1998a (for the HB), 
Ferraro, Bellazzini \& Fusi Pecci, 1995,
Ferraro et al. 1993b,  1997c, 1998b (for the Blue Stragglers).

This is the first in a series of papers devoted to study the
characteristics of the RGB, HB and AGB for the widest available sample of
GGCs with {\it good} $BV$ photometry.  In this paper we present the
catalog of the most recent CMDs for GGCs. From these we derive photometric
observables along RGB, HB, and AGB which yield new independent measures of
some peculiar features (e.g., the so-called RGB Bump).  The study will
soon be extended in a second paper to explore the existence and extent of
mixing processes (like the semiconvection and overshooting) in the stellar
interiors. These processes have a {\it direct} impact on the duration of
the post helium-flash phases (HB and AGB) and, in turn, on the use of 
population ratios to determine one of the fundamental cosmological
parameters, the helium abundance ($Y_p$), via the so-called {\it R-method}
(Iben 1968, Buzzoni et al 1983).

The paper is organized as follows: in Section 2, we present the complete 
data-base used in our analysis, which includes CMDs for 61 GGCs; while
Section 3 is devoted to the discussion of the metallicity scales.
Section 4 reports the basic assumptions of the theoretical
models adopted all along the paper.
In Section 5, we present a new procedure (based on the use of synthetic CMDs)
to measure the level of the Zero-Age HB (frequently adopted as {\it
standard luminosity reference}). Section 6 deals with the presentation
of new homogenous determinations of the RGB morphologic parameters, 
their calibrations in terms of the adopted spectroscopic and global 
metallicities, and the determination of the RGB-Bump luminosity and its 
comparison with the theoretical expectations. 
Similarly,
Section 7 is 
devoted to the study of the photometric
 properties of the AGB-Bump.
Finally, in Section 8, after
adopting different self-consistent distance scales, we report the results
of a global comparisons with the absolute quantities predicted by the 
theoretical models.

\section{THE DATABASE}

After reviewing the published literature on CMDs for GGCs, it is a little 
surprising to discover that the number of GGCs for which a modern (CCD) CMD is 
available is less than $50\%$ of the whole cluster population in the 
Galaxy. This percentage is further decreased if one restricts the sample,
as we did, to  only the clusters with CMDs of sufficient
photometric accuracy, population size, and degree of completeness down the
HB blue extension. 

Moreover, since we want to perform homogenous independent 
measures and star counts over the whole CMD, we  included in our
final sample only the GGCs whose data-sets (star magnitudes and positions) 
were available on electronic files (upon direct request to the author or 
scanned from the reference paper). In the final choice, we dropped CMDs
with inadequate completeness checks and usually adopted the most recent 
papers. If different CMDs of comparable quality 
were  available for the same cluster, after
carefully checking the radial extension of the samples and  their 
photometric compatibility, we merged them in order to 
increase the statistical significance of the adopted sample.

The final sample of GGCs whose CMDs have been classified as ``good
enough'' includes 61 objects, for which we list in Table 1: the name,
the metallicity in the considered scales (see next section), the
reddening from a recent compilation (Harris, 1996), and the
reference of the adopted CMD.

 Admittedly, the selected sample (listed in Table 1) is quite
heterogeneous in many respect: the cluster light sampling, the photometric
accuracy and the absolute calibration actually achieved by each
individual photometry.  However, no attempt has been performed at this
level to rank the clusters on the basis of the overall quality of the
CMD.  In the next paper, specifically devoted to presentation and
discussion of the population ratios, a more significative
classification will be performed on the basis of the global population
of bright stars (AGB + HB + RGB) sampled in each cluster.

\section{METALLICITY SCALES}

\subsection{The Zinn scale}

One of the most widely used scales for the metal abundance in GGCs has
been proposed during the early 80's by Zinn and his collaborators
(Zinn 1980, Zinn \& West 1984, Zinn 1985, hereafter Z85, and
Armandroff \& Zinn 1988). This scale was obtained from the integrated
light parameter $Q_{39}$ tied to the Cohen (1983) high dispersion and
low resolution spectrograms (see Zinn \& West 1984). Though 
dated, this metallicity scale is still the most complete (121 GGCs)
and homogeneous data-base available in the literature. In
the following, we will label as {Z85} the metallicity values
listed in column 4 of Table IV by Armandroff and Zinn (1988) or in
column 2 of Table I by Zinn (1985).
\subsection{The Carretta and Gratton scale}

Recently, Carretta \& Gratton (1997, hereafter CG97) have presented new 
measures of chemical abundances using high dispersion spectra for a set
of 24 GGCs, in the metallicity range ($-2.24<\feh_{\rm Z85}<-0.54$).
Though based on a small number of giants (the total sample includes $\sim 160$
stars, and in many cases only a few giants have been measured in each cluster),
these measures have the advantage of measuring {\it directly} the equivalent
widths of \fei\ and \feii\ lines. Comparing  their new  abundances to 
the Z85 metallicities, CG97 concluded that the Z85 scale is not linear and 
gave a quadratic relation suitable to transform the Z85 scale to their own 
scale (see eq. 7 in CG97).

As emphasized by CG97, the transformation relation 
can be safely used only in the metallicity range $-2.24<\feh_{\rm Z85}<-0.54$. 
In general, the CG97 scale turns out to yield  higher metallicity 
($\delta\feh\sim 0.2$) with respect to the Z85 scale for low-intermediate 
metallicity GGCs and  lower abundances ($\delta\feh\sim 0.1$) for metal 
rich GGCs.

For the sample of GGCs listed in Table 1 we eventually adopted the metallicity 
in the CG97 scale ($\feh_{\rm CG97}$) with the following assumptions:

\begin{enumerate}

\item 20 GGCs have direct spectroscopic measures in CG97.  For these clusters 
the value listed in Table 8 by CG97 has been adopted.

\item 35 GGCs in the quoted range of metallicity ($-2.24<\feh_{\rm Z85}<-0.54$)
have metallicities only in Z85. For them, we computed the $\feh_{\rm CG97}$ 
using eq. 7 of CG97.

\item 6 GGCs in our catalog (namely, NGC 5053, 5927, 6440, 6528,  6553, 
and Ter 7)
have Z85 values outside the validity range of the transformation to the CG97 
scale. For these objects we adopted $\feh_{\rm CG97} = 
\feh_{\rm Z85}+\delta\feh_{-0.54}$
and $\feh_{\rm CG97} = \feh_{\rm Z85}+\delta\feh_{-2.24}$
for clusters  with $\feh_{\rm Z85}>-0.54$
and $\feh_{\rm Z85}<-2.24$, respectively. Where $\delta\feh_{-0.54}$ and 
$\delta\feh_{-2.24}$ are the corrections, computed via eq. 7 by CG97, at 
$\feh_{\rm Z85}=-0.54$ and -2.24, respectively. 

\end{enumerate}

\subsection{Comparison with another recent catalog}

Recently Rutledge, Hesser \& Stetson (1997, hereafter RHS97) used homogeneous
observations of the CaII triplet lines in a sample of 71 GGCs in order
to measure an {\it abundance index}
which should provide a relatively accurate metallicity ranking.
 RHS97 calibrated this index in both the Zinn and CG97 metallicity scales.
Figure 1a,b shows the residuals between  the metallicities  
by RHS97 (in their Table 2) in the Z85 and CG97 scales, respectively,
and the values assumed in this paper
for the 42 clusters in common.
 As can be seen from both the panels, most of the clusters
are lying within $\pm 0.2$ dex (which is  a conservative but still
reliable level of the global accuracy 
for metal abundance determinations
for  GGCs). From Figure 1a,b it is evident that
the residuals do not show any  
trend with respect to the metallicity.
 Only a few clusters (namenly NGC5053, NGC6366 and Ter7 in
Fig1a and NGC5053 in Fig1b, respectively), show a larger 
($\delta[Fe/H]>0.4$) scatter and  deserve a more accurate 
spectroscopic analysis.

\clearpage
\hoffset = -10mm
\begin{deluxetable}{cccccc}
\tablewidth{18truecm}
\label{lm}
\tablecaption{The adopted data-base}
\tablehead{
\colhead{ $Name$} & 
\colhead{$\feh_{\rm Z85}$} & 
\colhead{$\feh_{\rm CG97}$} &
\colhead{ $\mh$} & 
\colhead{$E(B-V)$} & 
\colhead{$Reference$} }
\startdata
 NGC 104 &  -0.71 &  -0.70 &  -0.59 &   0.04 & Montegriffo et al.(1995)
 + Hesser et al (1987) \nl 
 NGC 288 &  -1.40 &  -1.07 &  -0.85 &   0.03 & 
 Bergbush (1993)+ Buonanno et al (1984)\nl
 NGC 362 &  -1.28 &  -1.15 &  -0.99 &   0.05 & Harris (1982) \nl 
 NGC1261 &  -1.31 &  -1.09 &  -0.89 &   0.02 & Ferraro et al. (1993a) \nl 
 NGC1466 &  -1.85 &  -1.64 &  -1.44 &   0.09 & Walker (1992a) \nl 
 NGC1841 &  -2.20 &  -2.11 &  -1.91 &   0.18 & Walker (1990) \nl
 NGC1851 &  -1.29 &  -1.08 &  -0.88 &   0.02 & Walker (1992b) \nl
 NGC1904 &  -1.69 &  -1.37 &  -1.22 &   0.01 & Ferraro et al. (1992) \nl 
 NGC2419 &  -2.10 &  -1.97 &  -1.77 &   0.03 & Christian et al. (1988) \nl 
 NGC2808 &  -1.37 &  -1.15 &  -0.95 &   0.23 & Ferraro et al. (1990) \nl 
 NGC3201 &  -1.61 &  -1.23 &  -1.03 &   0.21 & Covino et al. (1997) \nl 
 NGC4147 &  -1.80 &  -1.58 &  -1.38 &   0.02 & Sandage $\&$ Walker (1955) \nl 
 NGC4372 &  -2.08 &  -1.94 &  -1.74 &   0.45 & Brocato et al. (1996) \nl 
 NGC4590 &  -2.09 &  -1.99 &  -1.81 &   0.04 & Walker (1994) \nl 
 NGC4833 &  -1.86 &  -1.58 &  -1.27 &   0.33 & Momany (1996) \nl 
 NGC5053 &  -2.58 &  -2.51 &  -2.31 &   0.03 & Sarajedini \& Milone (1995) \nl 
 NGC5272 &  -1.66 &  -1.34 &  -1.16 &   0.01 &  Buonanno et al 1994+ 
Ferraro et al. (1997) \nl 
 NGC5286 &  -1.79 &  -1.57 &  -1.37 &   0.24 & Brocato et al. (1996)\nl
 NGC5466 &  -2.22 &  -2.14 &  -1.94 &   0.00 & Buonanno, Corsi $\&$ 
Fusi Pecci (1985) \nl 
 NGC5694 &  -1.91 &  -1.72 &  -1.52 &   0.09 & Ortolani $\&$ Gratton (1990) \nl 
 NGC5824 &  -1.85 &  -1.64 &  -1.44 &   0.14 & Bocato et al. (1996) \nl 
 NGC5897 &  -1.68 &  -1.59 &  -1.44 &   0.08 & Ferraro, Fusi Pecci $\&$ 
Buonanno (1992) \nl 
 NGC5904 &  -1.40 &  -1.11 &  -0.90 &   0.03 & Buonanno et al. (1981)+ 
Brocato et al. (1995) \nl 
 NGC5927 &  -0.31 &  -0.46 &  -0.37 &   0.47 & Samus et al. (1996) \nl 
 NGC6093 &  -1.64 &  -1.41 &  -1.21 &   0.18 & Brocato et al (1998) \nl 
 NGC6121 &  -1.33 &  -1.19 &  -0.94 &   0.36 & Lee (1977) \nl 
 NGC6171 &  -0.99 &  -0.87 &  -0.70 &   0.33 & Ferraro et al. (1991) \nl 
 NGC6205 &  -1.65 &  -1.39 &  -1.18 &   0.02 & Paltrinieri et al. (1998) \nl 
 NGC6218 &  -1.61 &  -1.37 &  -1.17 &   0.17 & Brocato et al. (1996) \nl 
 NGC6229 &  -1.54 &  -1.30 &  -1.10 &   0.01 & 
{\footnotesize 
Carney, Fullton, Trammell (1991)+ 
Borissova et al.(1997)} \nl 
 NGC6254 &  -1.60 &  -1.41 &  -1.25 &   0.28 & Harris, Racine ,De Roux (1976) \nl 
 NGC6266 &  -1.28 &  -1.07 &  -0.87 &   0.47 & Brocato et al. (1996) \nl 
 NGC6333 &  -1.78 &  -1.56 &  -1.36 &   0.36 & Janes $\&$ Heasley (1991) \nl 
 NGC6341 &  -2.24 &  -2.16 &  -1.95 &   0.02 & Buonanno, Corsi $\&$
Fusi Pecci (1985) \nl 
 NGC6352 &  -0.51 &  -0.64 &  -0.50 &   0.21 & Bordoni (1995) \nl 
 NGC6366 &  -0.99 &  -0.87 &  -0.70 &   0.69 & Pike (1976) \nl 
 NGC6397 &  -1.91 &  -1.82 &  -1.65 &   0.18 & Kaluzny (1997) \nl 
 NGC6440 &  -0.34 &  -0.49 &  -0.40 &   1.09 & Ortolani et al. (1994a) \nl 
 NGC6528 &  -0.23 &  -0.38 &  -0.31 &   0.62 & Ortolani et al. (1995) \nl 
 NGC6535 &  -1.75 &  -1.53 &  -1.33 &   0.32 & Sarajedini (1994a) \nl 
 NGC6553 &  -0.29 &  -0.44 &  -0.36 &   0.84 & Ortolani et al. (1995) \nl 
 NGC6584 &  -1.54 &  -1.30 &  -1.10 &   0.11 & Sarajedini $\&$ Forrester (1995) \nl 
 NGC6637 &  -0.59 &  -0.68 &  -0.55 &   0.17 & Ferraro et al. (1994) \nl 
 NGC6652 &  -0.99 &  -0.87 &  -0.70 &   0.09 & Ortolani et al. (1994b) \nl 
 NGC6681 &  -1.51 &  -1.27 &  -1.07 &   0.07 & Brocato et al. (1996) \nl 
 NGC6712 &  -1.01 &  -0.88 &  -0.71 &   0.46 & Cudworth (1988) \nl 
 NGC6717 &  -1.32 &  -1.10 &  -0.90 &   0.21 & Brocato et al. (1996) \nl 
 NGC6752 &  -1.54 &  -1.42 &  -1.21 &   0.04 & Buonanno et al. (1986) \nl 
 NGC6809 &  -1.82 &  -1.61 &  -1.41 &   0.07 & Desidera (1996) \nl 
 NGC6838 &  -0.58 &  -0.70 &  -0.49 &   0.25 & Cudworth (1995) \nl 
 NGC6934 &  -1.54 &  -1.30 &  -1.10 &   0.11 & Brocato et al. (1996) \nl 
 NGC6981 &  -1.54 &  -1.30 &  -1.10 &   0.05 & Brocato et al. (1996) \nl 
 NGC7006 &  -1.59 &  -1.35 &  -1.15 &   0.05 & Buonanno et al. (1991) \nl 
 NGC7078 &  -2.17 &  -2.12 &  -1.91 &   0.09 & Buonanno, Corsi $\&$ 
Fusi Pecci (1985) \nl 
 NGC7099 &  -2.13 &  -1.91 &  -1.71 &   0.03 & Bergbusch (1996) \nl 
 NGC7492 &  -1.51 &  -1.27 &  -1.07 &   0.00 & Buonanno et al. (1987) \nl 
 IC 4499 &  -1.50 &  -1.26 &  -1.06 &   0.24 & Ferraro et al. (1995) \nl 
 Rup 106 &  -1.90 &  -1.70 &  -1.50 &   0.21 & Buonanno et al. (1993) \nl 
 Arp   2 &  -1.85 &  -1.64 &  -1.44 &   0.11 & Buonanno et al. (1995a)\nl 
 Ter   7 &  -0.49 &  -0.64 &  -0.52 &   0.06 & Buonanno et al. (1995b) \nl 
 Ter   8 &  -1.81 &  -1.60 &  -1.40 &   0.14 & Ortolani $\&$ Gratton (1990) \nl 
\enddata
\end{deluxetable}

\hoffset = 0mm
 
\begin{figure}
\epsffile{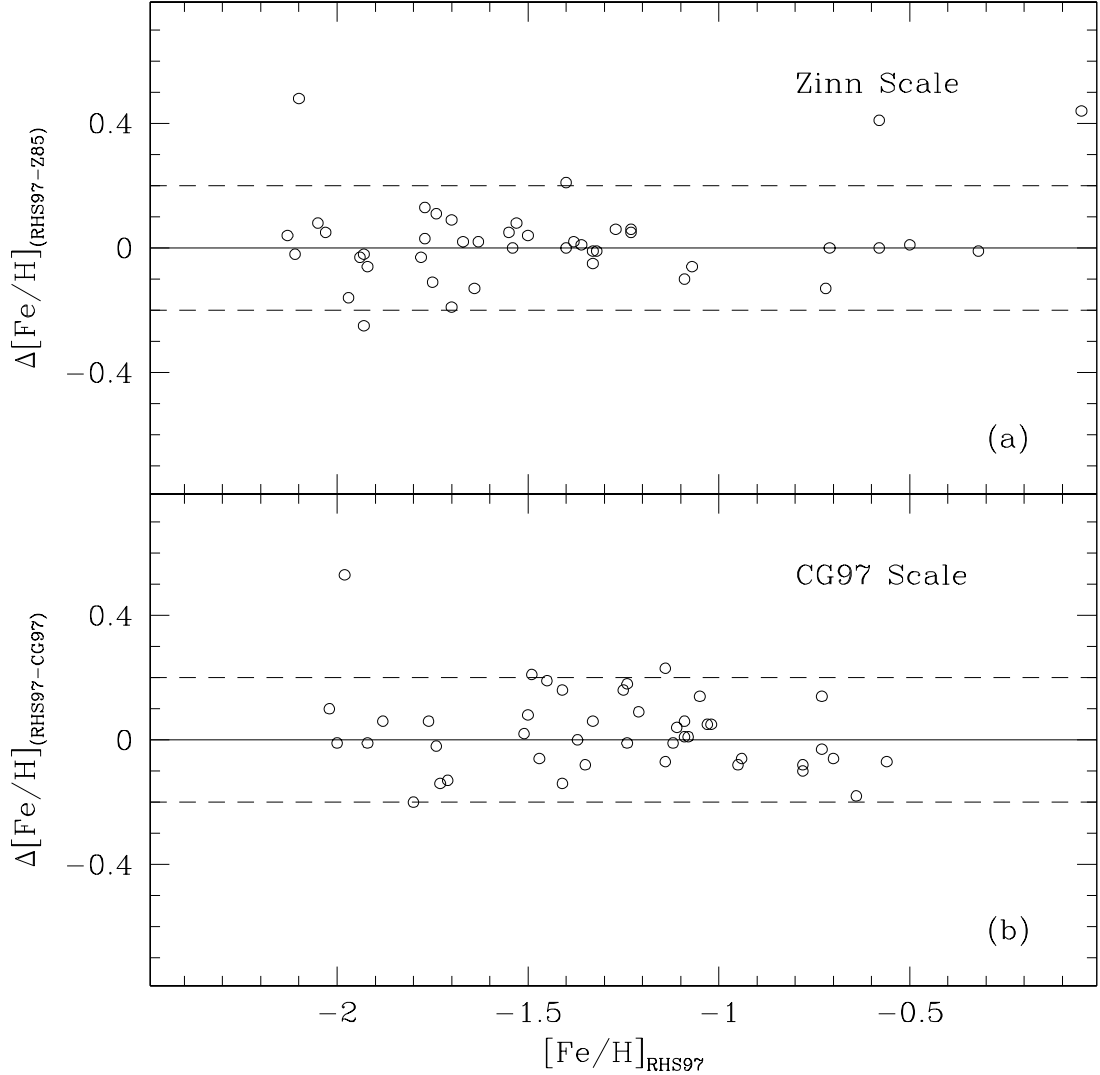}
\vspace{-2mm}
\caption{
\protect\label{map}
Residuals between metallicities by RHS97 (listed in their Table 2)  in the Zinn
  [{\it panel (a)}]  and in the CG97 scale [{\it panel (b)}], respectively, 
and the values assumed in this paper (see Table 1)
for the 42 clusters in common. 
}
\end{figure}

\begin{figure}
\epsffile{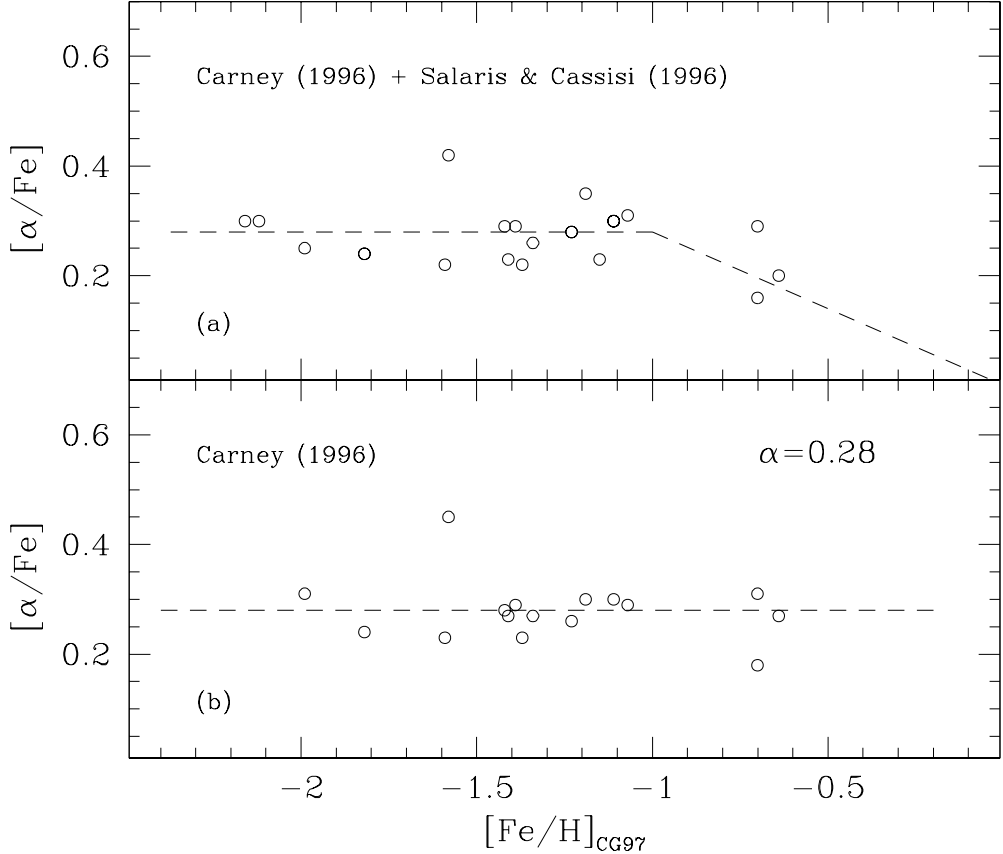}
\vspace{-2cm}
\caption{
\protect\label{map}
$[\alpha/Fe]$ as a function of $\feh_{\rm CG97}$.
{\it Panel (a)}: $[\alpha/Fe]$ are from  
 Salaris \& Cassisi (1996) and Carney (1996). For the 16 objects in common
in the two lists the $[\alpha/Fe]$ values have been averaged. 
 The dashed line is the {\it enhancement-relation}
 as a function of the metallicity which we assumed all along the paper.
{\it Panel (b)}: $[\alpha/Fe]$ measures are from Carney (1996).
The dashed line represents the scenario suggested by Carney (1996) for a
constant
 {\it enhancement} with  varying the metallicity.
}
\end{figure}

\begin{figure}
\epsffile{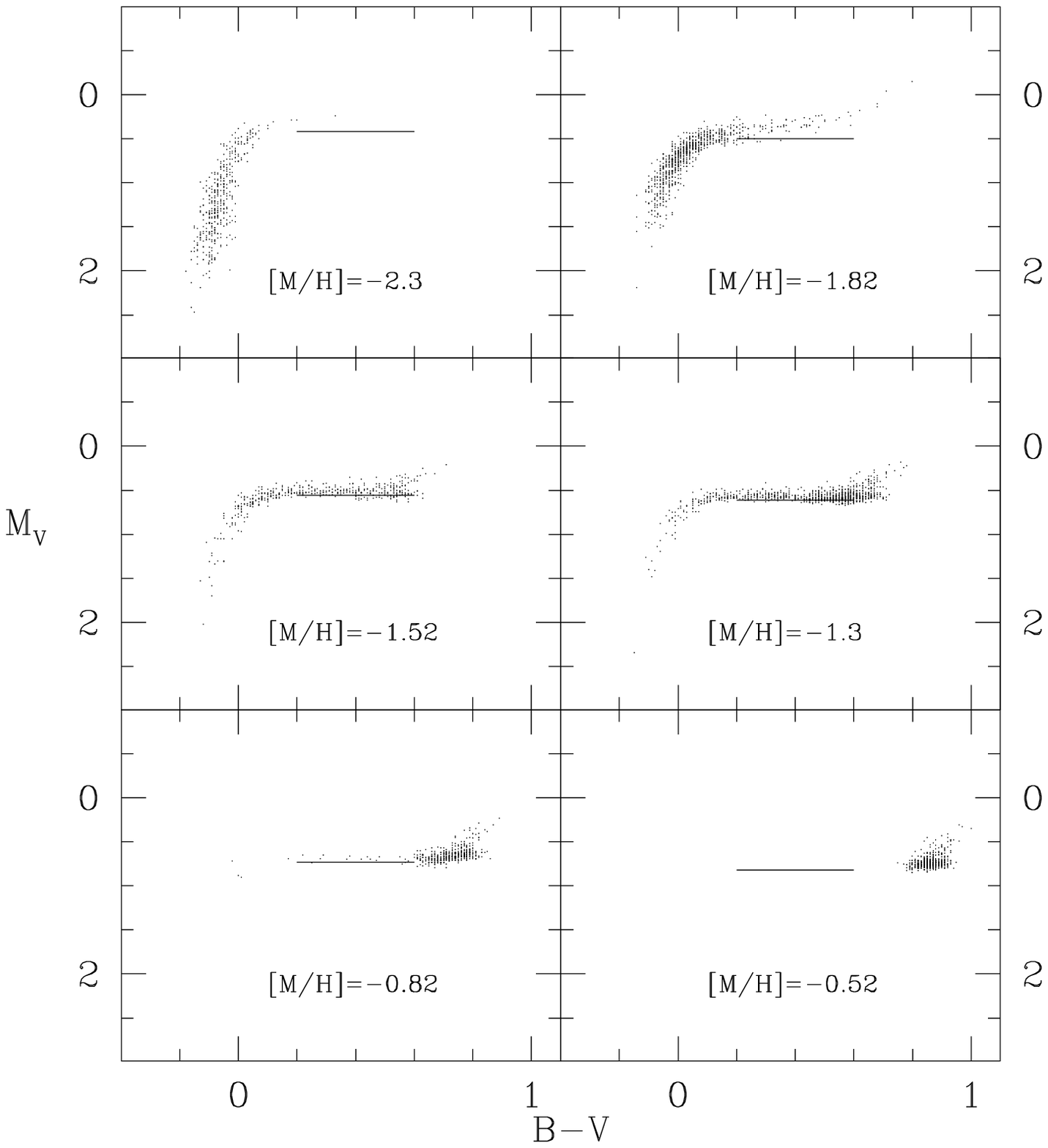}
\caption{
\protect\label{hbs}
Synthetic HB at different metallicities. The ZAHB level at 
$\logteff = 3.85$ is plotted as solid line.}
\end{figure}

These considerations strongly suggest that the two sets of measurements
are fully consistent within the global assumed uncertainty. 
In order to use the entire GGC data-set we collected,  
in the following discussion we adopt the metallicity values
(in the Zinn and CG97 scales) listed 
in Table 1. However, for sake of completeness,  
in Section 9  we further discuss 
the effect of adopting the metallicity  measurements
listed by RHS97 instead of those listed in Table 1.
 
\subsection{The global metallicity}

In the last decade it has become evident that in Population II stars
the abundance of $\alpha-$elements is enhanced with respect to iron.
Direct measurements of the $\alpha-$elements abundance in the halo field 
stars have shown a very well defined behaviour as a function of $\feh$, 
with a nearly constant overabundance ($\rm [\alpha/Fe]\sim0.4$) for $\feh<-2$ 
and a well defined trend with metallicity, which linearly decreases
to $\rm [\alpha/Fe]\sim0.0$  as metallicity increases (see Edvardsson et al 1993,
Nissen et al 1994, Magain 1989, Zhao \& Magain 1990, Gratton et al 1996).
In the GGC system the situation is not so clear. 
The mean overabundance seems to be $\rm [\alpha/Fe]\sim0.3$, but the behaviour 
with respect to the metallicity is still not firmly established. For example 
Carney (1996) claims that $\alpha-$element abundances do not appear to vary
as a function of \feh\ in GGCs. 

There are two recent compilations listing the $\alpha-$element
abundances measured in GGCs: Carney (1996) and Salaris \& Cassisi (1996).
Especially in the second list the data are collected from different
sources and are not the result of independent, self-consistent observations.
However, they can be used to  have useful quantitative hints.

In our catalog 16 GGCs  have values listed in Table 2 by Carney (1996), 
and 19 in Table 1 by Salaris \& Cassisi (1996). There are 16 objects in 
common in the two lists, and the values are in fairly good agreement (within 
$0.15$ dex). In the following,  we will adopt for $\rm [\alpha/Fe]$
the average of the values listed in the two tables. Figure 2a shows 
$\alphafe$ as a function of the metallicity in the CG97 scale.
Admittedly, it is hard to define a clear-cut trend with metallicity.
However,  the  $\alphafe$ abundance for $\feh_{\rm CG97}<-1$ is compatible
with a constant plateau, and for these clusters a mean value of $\alphafe=0.28$ 
has been adopted. At the metal rich extreme the situation is less clear.
There are only 3 clusters with $\feh_{\rm CG97}>-1$. The dashed 
line in Figure 2a
shows that their $\alpha-$element abundances are consistent with a
linear  decrease with increasing metallicity similar to thar seen in 
the field stars. For sake of simplicity, in the following
we have thus adopted such a trend
for the metal rich GGCs.
 However since the global trend of  $\rm [\alpha/Fe]$
with metallicity is still not firmly established, at least for GGCs,
 especially in the
high-metallicity domain,
we also considered the scenario suggested by Carney (1996), 
showed in Figure 2b. In this panel only the measures listed by Carney (1996)
 have been plotted. In Section 9, we show
which impact the use of the two proposed enhancement relations has on  
 on our results.

\begin{figure}
\epsffile{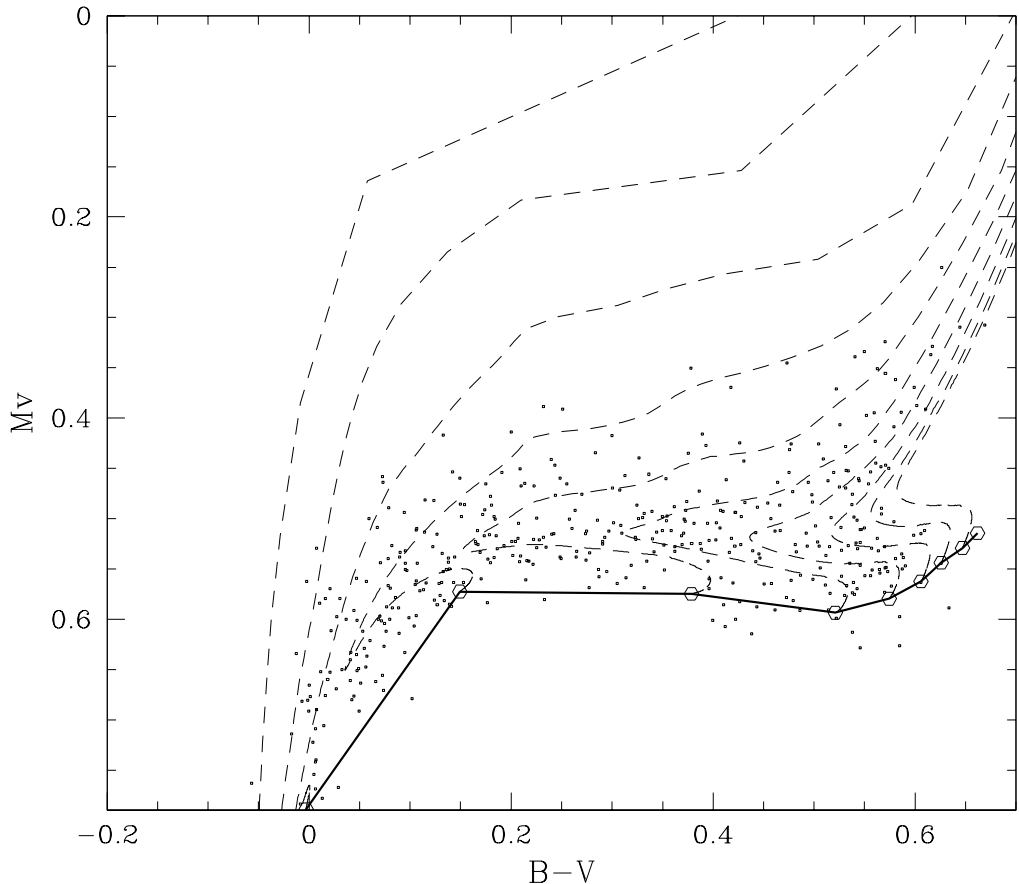}
\vspace{-2cm}
\caption{
\protect\label{tracce}
Evolutive tracks for $Log Z=-3.22$ are overplotted (as dashed lines)
to a synthetic HB. 
The heavy solid line is the ZAHB.}
\end{figure}

On the theoretical side, Salaris, Chieffi \& Straniero (1993) have 
investigated the effect produced on the theoretical evolutionary 
sequences by considering an enhancement of $\alpha-$elements.
They concluded that $\alpha-$enhanced isochrones are well
mimicked by the standard scaled-solar ones having global metallicity
\mh given by

\begin{equation}
\label{eq:2}
{\rm \mh} = {\rm \feh} + \log(0.638 f_{\alpha}+0.362)
\end{equation}
where $f_{\alpha}$ is the enhancement factor of the $\alpha-$elements.

Taking into account these prescriptions we computed the global
metallicity listed in column 4 of Table 1
as follows:

\begin{enumerate}

\item For the 19 GGCs with $\alphafe$  listed by 
 Salaris \& Cassisi (1996) or Carney (1996) we adopted  
$f_{\alpha}= 10^{\alphafe}$

\item For all the others,  we assumed
$f_{\alpha}= 10^{0.28}$ if $\feh<-0.8$
and 
$f_{\alpha}= 10^{-0.35\feh}$ if $\feh>-0.8$.

\end{enumerate}

\section{ MODELING THE RGB AND THE HB: THE STATE OF THE ART}

Understanding the observed properties of the HR-diagrams and
luminosity functions of GCCs stars necessarily requires an homogeneous
set of H and He-burning models of low mass stars and related
isochrones.  In this paper we have adopted the latest models computed by
using the FRANEC (Frascati Raphson Newton Evolutionary Code) first
described by Chieffi \& Straniero (1989). The  input physics has been
recently updated (see e.g. Straniero, Chieffi \& Limongi, 1997,
SCL97). A subset of these models have been presented in SCL97, while
the full set will be presented in a forthcoming paper (Chieffi,
Limongi \& Straniero 1998, hereafter CLS98).

The basic assumptions are here briefly summarized.

\begin{enumerate}

\item  The radiative opacity  coefficient is derived from 
the OPAL tables (Iglesias, Rogers
\& Wilson 1992) for temperatures larger then $10^4$ K, and from 
Alexander \& Fergusson (1994) at lower temperatures. 
Thermal  conduction is taken from
Itoh \etal\ (1983).

\item The equation  of  state  (EOS)  includes  quantum-relativistic  treatment
of the electron  plasma  (electron  degeneracy, pair production and the like). 
Coulomb  corrections  are  evaluated  by  means  of a Monte Carlo
technique using a revised 
version  of  the  Straniero (1988) 
EOS in which the partial degeneracy of the 
electron  component  is  taken  into  account  directly  in  the Monte Carlo 
calculations.

\item Thermal neutrinos rates due to plasma, 
photo and pair processes are taken into account following
the prescription of Munakata, Kohyama \& Itoh (1985), 
whereas Bremsstrahlung and recombination processes
are included following Dicus et al (1976) 
and Beaudet, Petrosian \& Salpeter (1967), respectively.  

\item Nuclear reaction rates are derived from Caughlan \& Fowler (1988), except
for the $^{12}{\rm C}(\alpha,\gamma)^{16}{\rm O}$ for which the rate of 
Caughlan et al. (1985) is used. 

\item The extension of the convective zones is determined by means of
the classical Schwarzschild criterium. Induced overshooting and
semiconvection during the central He-burning is also taken into
account (see Castellani et al. 1985). The mixing length theory is
adopted in order to evaluate the temperature gradient in the
convective regions.  The details of the mixing length calibrations can
be found in SCL97. 
Breathing pulses occurring at the end of the central He burning
phase have been inhibited  by adopting the procedure described in Caputo et
al (1989).

\item Microscopic diffusion of He and 
heavy elements have been included. Note that such a phenomenon 
mainly affects main sequence stars, while the  properties of post MS
evolution are only slightly changed (see SCL97 for more details). 

\end{enumerate}

Thus, models for masses ranging between 0.5 and 1.2\msun\ have been computed 
from the ZAMS up to the the onset of the He-flash. The range of metallicity 
covers the typical value of the GGCs, namely $0.0001{\le}Z{\le}0.006$. A constant 
He, $Y=0.23$, was adopted. Using these models we derived isochrones 
and luminosity functions for ages ranging between 8 and 20 Gyr.

For the HB, the present set includes models for masses ranging between
0.54 and 0.86\msun.  The same chemical compositions of the
corresponding H-burning models have been adopted. Following the usual
procedure, the core masses and the surface compositions of the ZAHB
models are derived from the corresponding last models of the H-burning
sequences. 
Their evolution
has been followed from the ZAHB up to the central He exhaustion.  Few
evolutionary sequences have been extended up to the first thermal
pulse on the AGB.
The procedure adopted to built the ZAHB is fully
described by Castellani \& Tornambe (1977). Breafly,
the first model in our HB sequence have a fully 
homogeneous H-rich envelope, but the ZAHB
model is set when all the secondary elements in the H-burning shell
are relaxed to their equilibrium values.
This happens when the zero age model has an age of $\sim 10^6$ yr.
Then, the age
of all the subsequent models have been rescaled to this zero point.

Finally, all the models have been transposed into the $V,~(B-V)$ plane
by means of the bolometric corrections and colour-temperature
relations obtained by Bessel, Castelli \& Pletz (1998a,b).

\section{THE OBSERVED ZAHB LEVEL: A NEW METHODOLOGICAL APPROACH}

Since the first wide series of HB models (Rood 1973), it is well known that
the observed HB cannot be described by any  single evolutionary track.
It can rather be modelled by convolving a proper set of evolutionary paths
of stars having slightly different values of total and/or core masses 
(Rood 1973). In other words, one can imagine the so-called Zero Age HB
(ZAHB)  as a sort of {\it starting locus} where stars are located after 
the helium ignition in the core, depending on their total and core mass, 
and from which they start their evolutionary run toward the AGB. 
It is thus quite simple (at least in principle) to accurately determine 
the location of the theoretical ZAHB.

On the observational side, measuring the ZAHB level is unfortunately
fairly difficult and sometimes ambiguous.  To minimize any possible
evolutionary effects off the ZAHB, one might ideally define the ZAHB
level by adopting the magnitude of the lower envelope of the observed
HB distribution in the region with $0.2<(B-V)<0.6$.  

However, the ``HB
levels'' found in the literature are most often not compatible each
other (and not directly comparable) as they actually are the mean
level of the HB ($<V_{\rm HB}>$), or the mean magnitude of the RR
Lyrae stars ($<V_{\rm RR}>$), or, finally, the ``estimated'' ZAHB
level.  Indeed, the frequent (implicit) assumption that $<V_{\rm RR}>$
is coincident with the ZAHB level has been largely criticized (see for
example Lee, Demarque \& Zinn, 1990) as the actual difference between
these two quantities strongly depends on the HB morphology and, in
turn, on metallicity (see Carney, Storm \& Jones, 1992 and Cassisi \&
Salaris 1997).

\clearpage
\hoffset = -10mm
\begin{deluxetable}{cccccccl}
\tablewidth{17.5truecm}
\label{lm}
\tablecaption{$V_{ZAHB}$, metallicities, reddening and derived DMs for the 
program GGCs.}
\tablehead{
\colhead{$Name$} & 
\colhead{$\feh_{\rm Z85}$} & 
\colhead{$\feh_{\rm CG97}$} & 
\colhead{$\mh$} & 
\colhead{$E(B-V)$} & 
\colhead{$V_{ZAHB}$} &
\colhead{$(M-m)_0^{CG97}$} & 
\colhead{$(M-m)_0^{\mh}$}
}
\startdata
 NGC 104 &  -0.71 &  -0.70 &  -0.59 &   0.04 &  14.22$\pm$0.07 &  13.32 &  13.29 \nl 
 NGC 288 &  -1.40 &  -1.07 &  -0.85 &   0.03 &  15.50$\pm$0.10 &  14.73 &  14.67 \nl
 NGC 362 &  -1.28 &  -1.15 &  -0.99 &   0.05 &  15.50$\pm$0.07 &  14.68 &  14.64 \nl
 NGC1261 &  -1.31 &  -1.09 &  -0.89 &   0.02 &  16.72$\pm$0.05 &  15.98 &  15.93 \nl
 NGC1466 &  -1.85 &  -1.64 &  -1.44 &   0.09 &  19.30$\pm$0.07 &  18.47 &  18.43 \nl
 NGC1841 &  -2.20 &  -2.11 &  -1.91 &   0.18 &  19.42$\pm$0.10 &  18.39 &  18.36 \nl
 NGC1851 &  -1.29 &  -1.08 &  -0.88 &   0.02 &  16.20$\pm$0.05 &  15.46 &  15.41 \nl
 NGC1904 &  -1.69 &  -1.37 &  -1.22 &   0.01 &  16.27$\pm$0.07 &  15.63 &  15.60 \nl
 NGC2419 &  -2.10 &  -1.97 &  -1.77 &   0.03 &  20.50$\pm$0.10 &  19.92 &  19.88 \nl
 NGC2808 &  -1.37 &  -1.15 &  -0.95 &   0.23 &  16.27$\pm$0.07 &  14.90 &  14.85 \nl
 NGC3201 &  -1.61 &  -1.23 &  -1.03 &   0.21 &  14.77$\pm$0.07 &  13.48 &  13.43 \nl
 NGC4147 &  -1.80 &  -1.58 &  -1.38 &   0.02 &  16.95$\pm$0.10 &  16.32 &  16.28 \nl
 NGC4372 &  -2.08 &  -1.94 &  -1.74 &   0.45 &  15.90$\pm$0.15 &  14.01 &  13.97 \nl
 NGC4590 &  -2.09 &  -1.99 &  -1.81 &   0.04 &  15.75$\pm$0.05 &  15.14 &  15.11 \nl
 NGC4833 &  -1.86 &  -1.58 &  -1.27 &   0.33 &  15.77$\pm$0.07 &  14.18 &  14.12 \nl
 NGC5053 &  -2.58 &  -2.51 &  -2.31 &   0.03 &  16.70$\pm$0.07 &  16.19 &  16.17 \nl
 NGC5272 &  -1.66 &  -1.34 &  -1.16 &   0.01 &  15.68$\pm$0.05 &  15.03 &  14.99 \nl
 NGC5286 &  -1.79 &  -1.57 &  -1.37 &   0.24 &  16.60$\pm$0.10 &  15.29 &  15.25 \nl
 NGC5466 &  -2.22 &  -2.14 &  -1.94 &   0.00 &  16.62$\pm$0.10 &  16.16 &  16.12 \nl
 NGC5694 &  -1.91 &  -1.72 &  -1.52 &   0.09 &  18.70$\pm$0.10 &  17.88 &  17.84 \nl
 NGC5824 &  -1.85 &  -1.64 &  -1.44 &   0.14 &  18.52$\pm$0.07 &  17.53 &  17.49 \nl
 NGC5897 &  -1.68 &  -1.59 &  -1.44 &   0.08 &  16.45$\pm$0.07 &  15.64 &  15.61 \nl
 NGC5904 &  -1.40 &  -1.11 &  -0.90 &   0.03 &  15.13$\pm$0.05 &  14.37 &  14.31 \nl
 NGC5927 &  -0.31 &  -0.46 &  -0.37 &   0.47 &  16.72$\pm$0.10 &  14.41 &  14.39 \nl
 NGC6093 &  -1.64 &  -1.41 &  -1.21 &   0.18 &  16.12$\pm$0.07 &  14.96 &  14.92 \nl
 NGC6121 &  -1.33 &  -1.19 &  -0.94 &   0.36 &  13.45$\pm$0.10 &  11.68 &  11.62 \nl
 NGC6171 &  -0.99 &  -0.87 &  -0.70 &   0.33 &  15.70$\pm$0.10 &  13.95 &  13.90 \nl
 NGC6205 &  -1.65 &  -1.39 &  -1.18 &   0.02 &  15.10$\pm$0.15 &  14.43 &  14.38 \nl
 NGC6218 &  -1.61 &  -1.37 &  -1.17 &   0.17 &  14.75$\pm$0.15 &  13.61 &  13.57 \nl
 NGC6229 &  -1.54 &  -1.30 &  -1.10 &   0.01 &  18.11$\pm$0.05 &  17.45 &  17.41 \nl
 NGC6254 &  -1.60 &  -1.41 &  -1.25 &   0.28 &  14.85$\pm$0.10 &  13.38 &  13.35 \nl
 NGC6266 &  -1.28 &  -1.07 &  -0.87 &   0.47 &  16.40$\pm$0.20 &  14.26 &  14.21 \nl
 NGC6333 &  -1.78 &  -1.56 &  -1.36 &   0.36 &  16.35$\pm$0.15 &  14.67 &  14.62 \nl
 NGC6341 &  -2.24 &  -2.16 &  -1.95 &   0.02 &  15.30$\pm$0.10 &  14.78 &  14.74 \nl
 NGC6352 &  -0.51 &  -0.64 &  -0.50 &   0.21 &  15.30$\pm$0.10 &  13.85 &  13.81 \nl
 NGC6366 &  -0.99 &  -0.87 &  -0.70 &   0.69 &  15.80$\pm$0.10 &  12.93 &  12.88 \nl
 NGC6397 &  -1.91 &  -1.82 &  -1.65 &   0.18 &  13.00$\pm$0.10 &  11.92 &  11.89 \nl
 NGC6440 &  -0.34 &  -0.49 &  -0.40 &   1.09 &  18.70$\pm$0.20 &  14.48 &  14.45 \nl
 NGC6528 &  -0.23 &  -0.38 &  -0.31 &   0.62 &  17.17$\pm$0.20 &  14.37 &  14.35 \nl
 NGC6535 &  -1.75 &  -1.53 &  -1.33 &   0.32 &  15.90$\pm$0.15 &  14.33 &  14.29 \nl
 NGC6553 &  -0.29 &  -0.44 &  -0.36 &   0.84 &  16.92$\pm$0.20 &  13.46 &  13.44 \nl
 NGC6584 &  -1.54 &  -1.30 &  -1.10 &   0.11 &  16.60$\pm$0.05 &  15.63 &  15.59 \nl
 NGC6637 &  -0.59 &  -0.68 &  -0.55 &   0.17 &  15.95$\pm$0.10 &  14.64 &  14.60 \nl
 NGC6652 &  -0.99 &  -0.87 &  -0.70 &   0.09 &  16.07$\pm$0.10 &  15.06 &  15.01 \nl
 NGC6681 &  -1.51 &  -1.27 &  -1.07 &   0.07 &  15.85$\pm$0.10 &  15.00 &  14.95 \nl
 NGC6712 &  -1.01 &  -0.88 &  -0.71 &   0.46 &  16.32$\pm$0.07 &  14.16 &  14.12 \nl
 NGC6717 &  -1.32 &  -1.10 &  -0.90 &   0.21 &  15.75$\pm$0.15 &  14.43 &  14.38 \nl
 NGC6752 &  -1.54 &  -1.42 &  -1.21 &   0.04 &  13.90$\pm$0.15 &  13.18 &  13.13 \nl
 NGC6809 &  -1.82 &  -1.61 &  -1.41 &   0.07 &  14.60$\pm$0.10 &  13.82 &  13.78 \nl
 NGC6838 &  -0.58 &  -0.70 &  -0.49 &   0.25 &  14.52$\pm$0.10 &  12.97 &  12.90 \nl
 NGC6934 &  -1.54 &  -1.30 &  -1.10 &   0.11 &  16.97$\pm$0.07 &  16.00 &  15.96 \nl
 NGC6981 &  -1.54 &  -1.30 &  -1.10 &   0.05 &  16.86$\pm$0.07 &  16.08 &  16.03 \nl
 NGC7006 &  -1.59 &  -1.35 &  -1.15 &   0.05 &  18.85$\pm$0.15 &  18.08 &  18.03 \nl
 NGC7078 &  -2.17 &  -2.12 &  -1.91 &   0.09 &  15.90$\pm$0.07 &  15.15 &  15.12 \nl
 NGC7099 &  -2.13 &  -1.91 &  -1.71 &   0.03 &  15.30$\pm$0.10 &  14.71 &  14.67 \nl
 NGC7492 &  -1.51 &  -1.27 &  -1.07 &   0.00 &  17.78$\pm$0.10 &  17.15 &  17.10 \nl
 IC4499 &  -1.50 &  -1.26 &  -1.06 &   0.24 &  17.70$\pm$0.07 &  16.32 &  16.27 \nl
 Rup 106 &  -1.90 &  -1.70 &  -1.50 &   0.21 &  17.85$\pm$0.10 &  16.66 &  16.62 \nl 
 Arp   2 &  -1.85 &  -1.64 &  -1.44 &   0.11 &  18.30$\pm$0.15 &  17.41 &  17.37 \nl
 Ter   7 &  -0.49 &  -0.64 &  -0.52 &   0.06 &  17.87$\pm$0.10 &  16.89 &  16.85 \nl
 Ter   8 &  -1.81 &  -1.60 &  -1.40 &   0.14 &  18.15$\pm$0.10 &  17.16 &  17.11 \nl
\enddata
\end{deluxetable}

\hoffset = 0mm
 
To overcome these ambiguities,  we have developed a new procedure 
to re-determine the ZAHB level for all the GGCs listed in our catalog,
so as to yield values which could be homogeneous and directly comparable
with the corresponding theoretical ones. 
Using the full set of HB evolutionary tracks described in the previous
section we generated a wide sample of synthetic HBs. 
The method and the code adopted to derive the synthetic HR-diagrams are described 
in a forthcoming paper (Chieffi, Straniero \& Limongi 1998, in preparation). 
Briefly, to model an observed HB of a cluster with a given chemical
composition this code requires several input parameters: 

\begin{description}

\item[$V_{lim}$---]  the photometric limiting magnitude of the synthetic CMD

\item[$N_{\rm HB}$---]  the total number of stars with $V<V_{\rm lim}$

\item[$M_{\rm HB}$---] the mean mass of the HB stars (which drives the position 
in colour of the bulk of the star distribution along the ZAHB)

\item[$\sigma_M$---] the width of the gaussian mass distribution (which drives the 
spread in colour of the HB stars).

\item[photometric errors---] When comparing the synthetic HR-diagrams
with the real ones, we have also to specify the photometric error bar
the completeness of the stellar sample at different luminosity.

\end{description}

By properly tuning these quantities it is possible, in principle, to
reproduce any observed HB morphology. For example, $\sigma_M$ is the
main parameter driving the presence and the extension of the HB blue
tail. For example, in Figure 3 we show a set of synthetic HBs
for 6 prototype clusters with different metallicity and, in turn,
different HB morphologies. The horizontal line indicates the 
level of the ZAHB at $\logteff= 3.85$.  As already noted,  the lower 
envelope of the star distribution is not always coincident  with the 
ZAHB level at $\logteff= 3.85$, and this confirms the need of a careful
procedure to yield meaningful and comparable values for the ZAHB.

The problem is illustrated  in a more appropriate scale in Figure 4 which 
shows  the evolution of the HB stars off the ZAHB level.
From inspecting this figure it is evident that the ZAHB level is not 
coincident with the lower envelope of the HB star distribution even 
when the HB is uniformly populated in the RR Lyrae region.
This effect is mainly due to the fact that the evolution away 
from the ZAHB is quite rapid at the beginning. 
After only 8 Myr ($\sim 8\%$ of the 
total lifetime in HB) the stars are already 0.05--0.1 mag brighter than the 
ZAHB {\it starting line}, then they spend $\sim 70\%$ of the total HB 
time in covering the next 0.1 mag. Thus the near-ZAHB HB is inherently
poorly populated and the observed lower envelope of the HB will be a
poor measure of the ``HB level'' which is affected both by the sample
size and the size of the photometric errors.

\begin{figure}
\epsffile{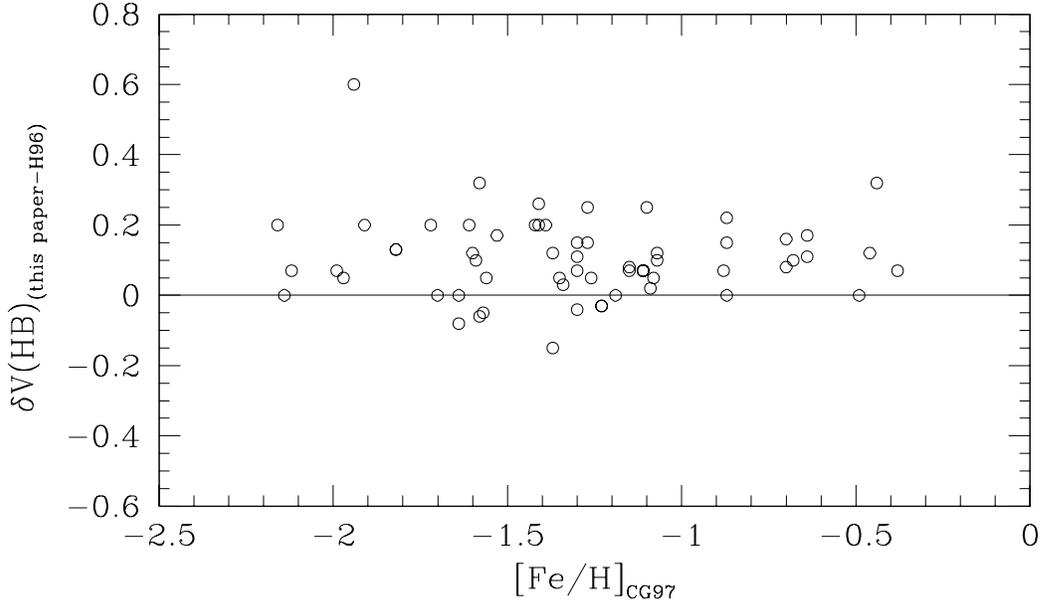}
\vspace{-3cm}
\caption{
\protect\label{vhb}
Differences between the $V_{ZAHB}$ level obtained in this paper and
the $V(HB)$ listed by H96.
}
\end{figure}

\begin{figure}
\epsffile{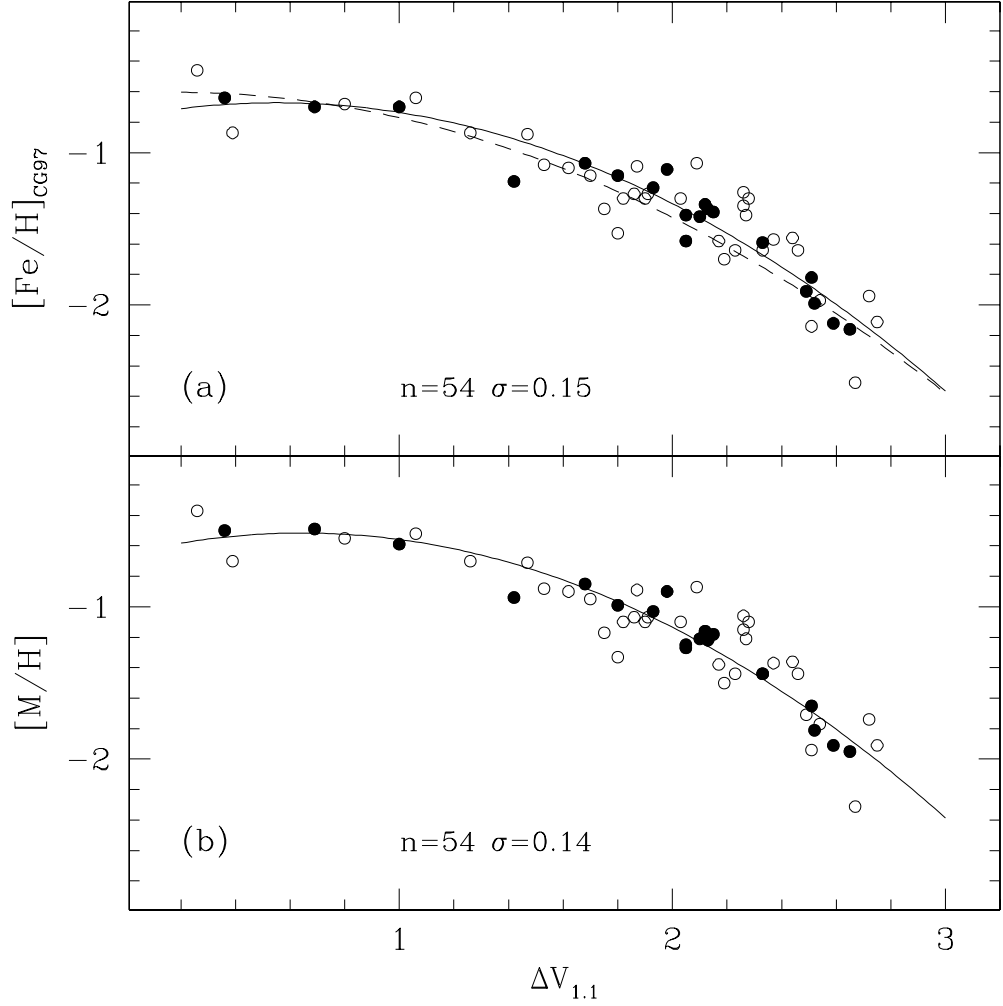}
\caption{
\protect\label{rgb1}
Calibration of the $\Delta V_{1.1}$ parameter with respect $\feh_{\rm CG97}$
({\it panel (a)}) and the global metallicity ($\mh$) ({\it panel (b)}).
The filled symbols represent clusters for which spectroscopic
metallicity and $\alphafe$ abundance has been directly measured. The solid
lines are the best fit to the data. The dashed line in {\it panel (a)}
is the relation recently obtained by Carretta \& Bragaglia 1998. 
The number of clusters used to compute each relation is reported together
with the standard deviations of the data.
}
\end{figure}

\begin{figure}
\epsffile{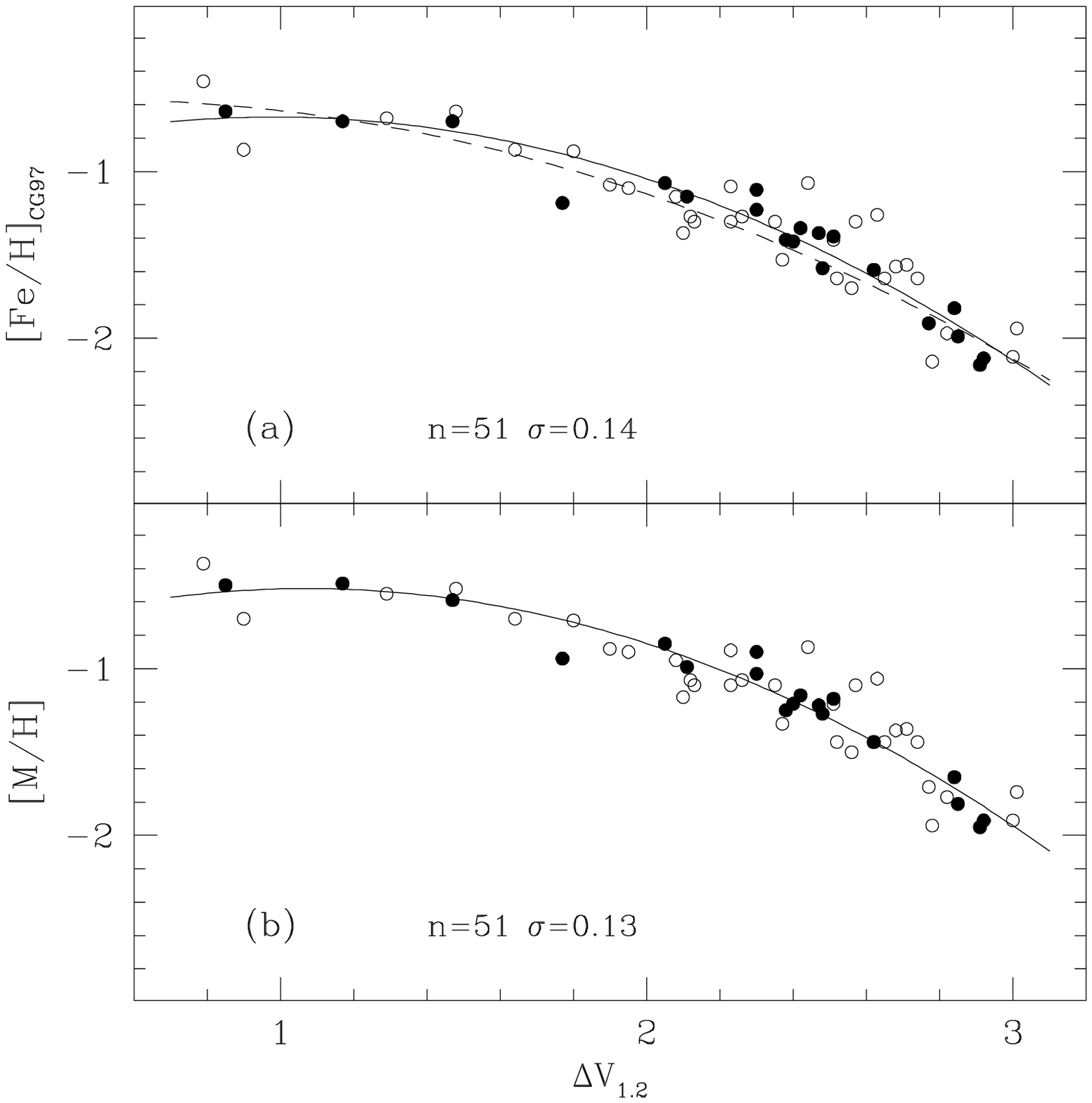}
\caption{
\protect\label{rgb2}
Calibration of the $\Delta V_{1.2}$ parameter with respect $\feh_{\rm CG97}$
({\it panel (a)}) and the global metallicity ($\mh$) ({\it panel (b)}).
The symbols have the same meaning of Figure 6.
}
\end{figure}

\begin{figure}
\epsffile{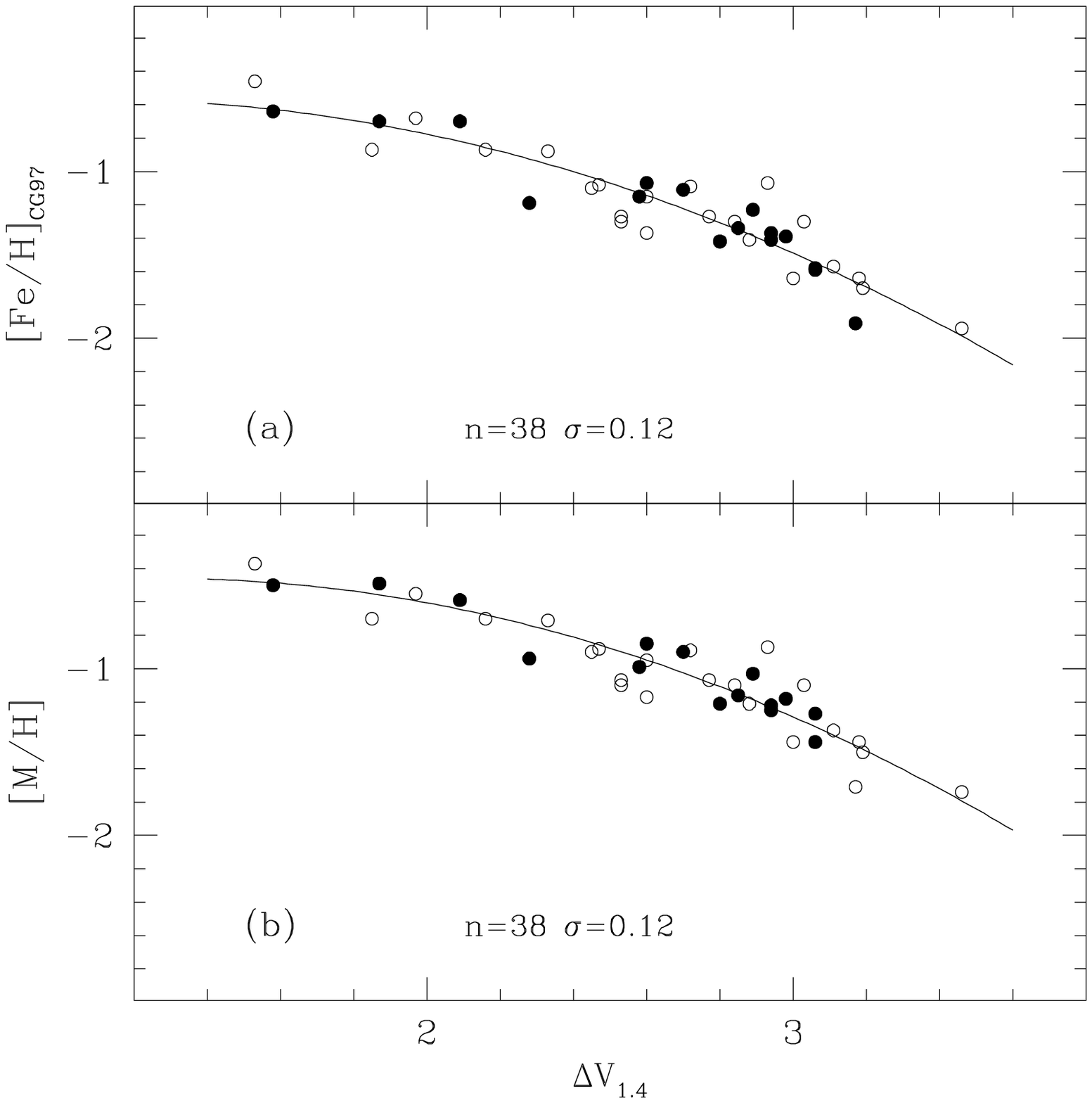}
\caption{
\protect\label{rgb3}
Calibration of the $\Delta V_{1.4}$ parameter with respect $\feh_{\rm CG97}$
({\it panel (a)}) and the global metallicity ($\mh$) ({\it panel (b)}).
The symbols have the same meaning of Figure 6.
}
\end{figure}

To overcome the problem we have  adopted the following 
empirical procedure:

\begin{enumerate}

\item For each cluster in our catalog we computed a synthetic HB (with the
appropriate abundances) tuning the parameters so as to best
reproduce the observed HB morphology.

\item The synthetic HB has been shifted in magnitude and colour to
match the observed HB.

\item The ``observed'' $V_{ZAHB}$ value is then read from the line
indicating the theoretical ZAHB level at $\logteff= 3.85$ as yielded
by the best-fitting synthetic HB.

\end{enumerate}

Note that this procedure does makes use of the models only as a guide
to drawing of the location of the ``true'' ZAHB level and is only
slightly dependent on the adopted theoretical models. It avoids the
uncertainties induced by the differences in the observed HB
morphologies and yields values obtained with a homogeneous and
self-consistent empirical method. ZAHB levels determined in this way
should be especially appropriate to compare with theoretical
models. The $V_{\rm ZAHB}$ value thus obtained are listed in column 6
of Table 2.  The errors in $V_{\rm ZAHB}$ has been estimated by
combining the scatter from multiple independent determinations of the
ZAHB level and an estimate of the photometric error (at the HB level)
in each individual cluster.

Although the actual difference between $V_{\rm ZAHB}$
and  $<V_{HB}>$ depends on the various parameters
(like the mean star mass, the core mass, the metallicity,
 helium abundance, etc) which drive the HB morphology, 
on the base of the synthetic HB plotted in Figure 3 and Figure 4  
 we derived the following average relation:

\begin{equation}
\label{eq:1}
{\rm V_{\rm ZAHB}} = <V_{HB}> +0.106 \mh^{2} +0.236 \mh +0.193\\
\end{equation}

which can be used, at a first order,
 to derive the $V_{\rm ZAHB}$ level from the
$<V_{HB}> $ measured in the colour range $0.2<(B-V)<0.6$. Note that
for metal rich clusters ($[M/H]>-1$) the mean value of the
red HB clump  was assumed as $<V_{HB}> $. The relation suggests 
that the minimum
difference between ZAHB and mean HB luminosity ($\delta V\sim 0.06$)
occurs at $[M/H]\sim -1.2$, and it turns to be $\delta V\sim 0.16$
and $\sim 0.10$ at $[M/H]=-2.2$ and $-0.5$, respectively.

We can compare the adopted $V_{\rm ZAHB}$
values listed in Table 2, for instance, with those listed in
the recent compilation by Harris (1996)(hereafter H96).  The residuals
($\rm this~paper-H96$) are plotted versus $\feh_{\rm CG97}$ in Figure
5. As expected, there is a clear systematic difference ($\sim 0.17$
mag) between the two, with the values derived in this paper being
fainter than those listed by H96.  Only one cluster (NGC 4372) shows a
large ($\delta V>0.4$) residual. This is due to the fact that H96
adopted a different (older) photometry (with a different photometric
zero-point) than that used in this paper.  Similar comparisons can be
made with other compilations (Buonanno, Corsi \& Fusi Pecci 1989,
Chaboyer, Demarque \& Sarajedini, 1996). In both cases $V_{\rm
HB}$ values listed in Table 2 are systematically fainter ($\delta V
\sim 0.1$ and $\delta V \sim 0.15$, respectively).

\section{ THE RED GIANT BRANCH}

\subsection {The RGB mean ridge line}

In order to derive the mean ridge line of the RGB for all the GGCs listed in 
our catalog we adopted the following procedure:

\begin{enumerate}

\item  A rough preliminary  selection of the stars belonging to the RGB
(excluding the HB and AGB stars) has been performed by eye to initialize
and to accelerate the subsequent iterations.

\item The polynomial fitting technique presented by Sarajedini \&
Norris (1994, hereafter SN94) has then be applied to the samples. In
particular, the RGB has been fitted by a ($\rm 2^{nd} ~{or}~3^{rd}$ order)
polynomial law in the form $(B-V)=f(V)$. After each iteration stars
more than 2-$\sigma$ in colour from the best-fitting ridge line were rejected
and the fitting procedure repeated to yield a stable solution.
\end{enumerate}

\subsection {Photometric parameters along the RGB}

As widely known, RGB morphology and location in the CMD are good metallicity 
indicators of the parent cluster. In particular three main parameters 
have been defined to describe the photometric characteristics of the RGB:

\begin{description}

\item[$\Delta V$ --] First defined by Sandage and Wallerstein (1960),
as a measure of the height of the RGB brighter than the HB level. They
used $\Delta V_{1.4}$ (in mag.), with $V_{\rm RGB}$ measured at the intrinsic colour
$(B-V)_0=1.4$.  Recently, Sarajedini \& Layden (1997, SL97) have
defined two similar parameters, $\Delta V_{1.1}$ and $\Delta V_{1.2}$,
measured at $(B-V)_0=1.1$ and $(B-V)_0=1.2$, respectively. These two
additional parameters are particularly useful since the
observed samples are often not populated enough at $(B-V)_0=1.4$ to clearly
define a mean ridge line.

\item[$(B-V)_{0,g}$ --] Defined by Sandage and Smith (1966) as the intrinsic 
colour of the RGB at the HB level.

\item[$S$ --] Defined by Hartwick (1968) (and here called
$S_{2.5}$) as the slope of the line connecting two points along the
RGB: the first being intersection of the RGB with the line defining
the HB level and the second being the point on the RGB 2.5 mag
brighter than the HB. Following SL97, we have also defined $S_{2.0}$
which is based on the the RGB point only 2.0 mag brighter than the HB
level.

\end{description}

\begin{figure}
\epsffile{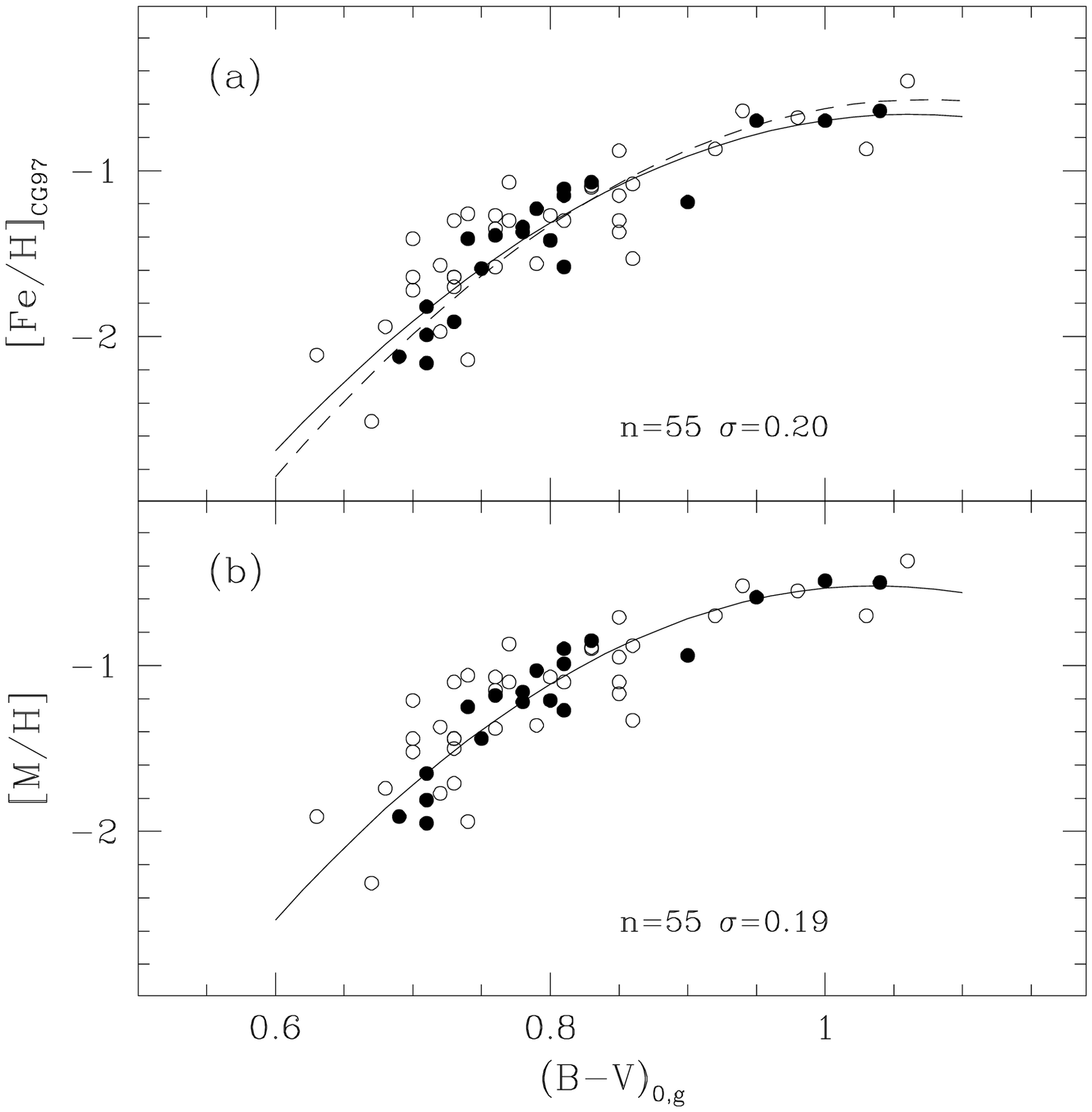}
\caption{
\protect\label{rgb4}
Calibration of the $(B-V)_{0,g}$ parameter with respect $\feh_{\rm CG97}$
({\it panel (a)}) and the global metallicity ($\mh$) ({\it panel (b)}).
The symbols have the same meaning of Figure 6.
}
\end{figure}

\begin{figure}
\epsffile{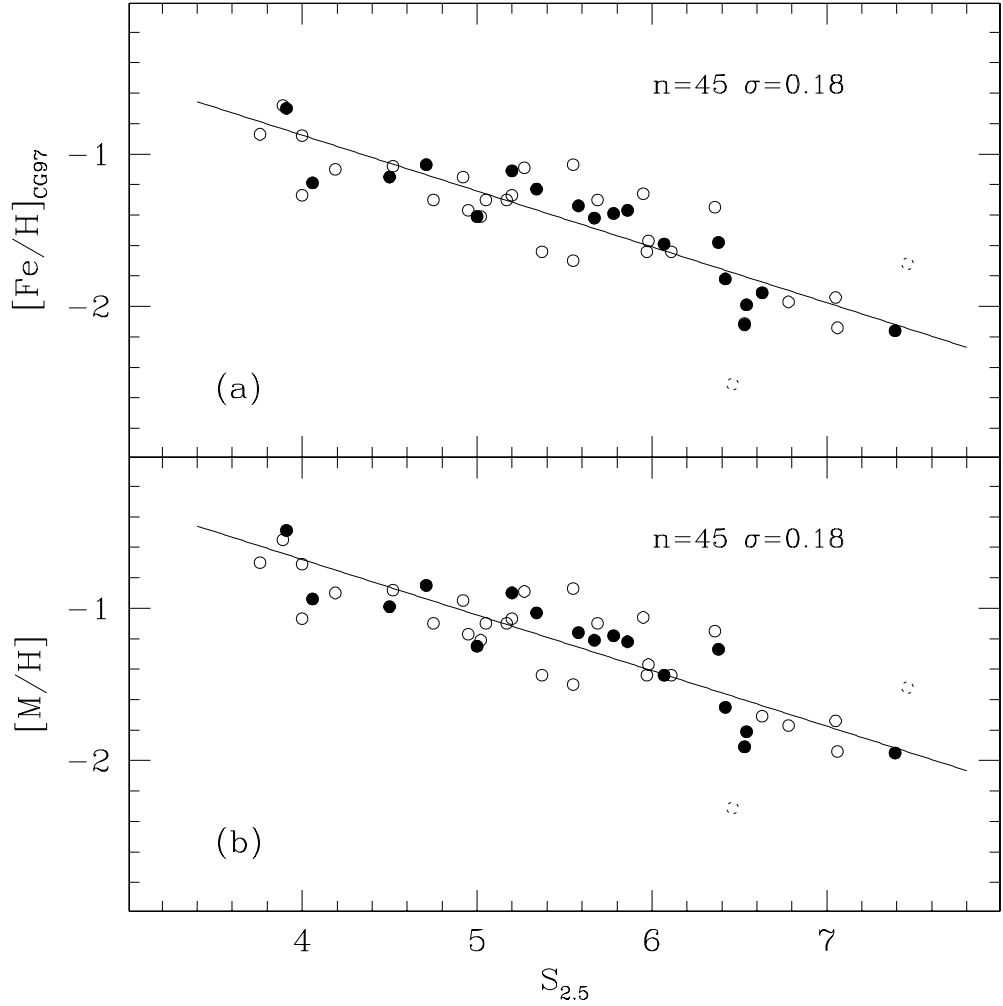}
\caption{
\protect\label{rgb5}
Calibration of the $S_{2.5}$ parameter with respect $\feh_{\rm CG97}$
({\it panel (a)}) and the global metallicity ($\mh$) ({\it panel (b)}).
The symbols have the same meaning of Figure 6.
Two clusters 
(namely NGC 5053 and NGC 5694,
plotted as dotted circles), 
have been excluded in the determination of 
the fitting relation (see text).
}
\end{figure}

\begin{figure}
\epsffile{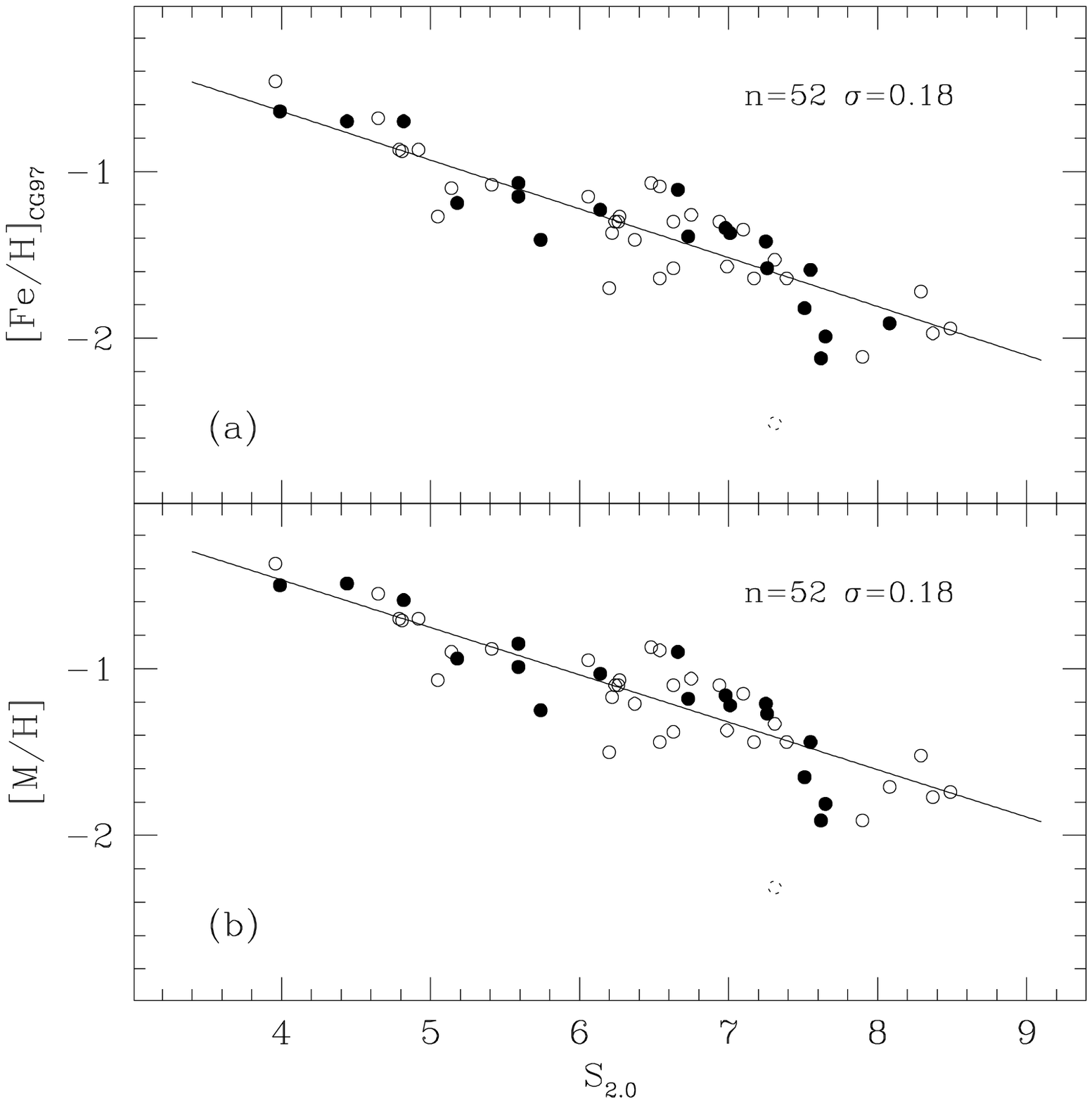}
\caption{
\protect\label{rgb6}
Calibration of the $S_{2.0}$ parameter with respect $\feh_{\rm CG97}$
({\it panel (a)}) and the global metallicity ($\mh$) ({\it panel (b)}).
The symbols have the same meaning of Figure 6.
}
\end{figure}

\begin{deluxetable}{cccccccccc}
\tablewidth{17truecm}
\tablecaption{RGB  parameters.}
\tablehead{
\colhead{ $Name$} & 
\colhead{[Fe/H]$_{CG97}$} & 
\colhead{[M/H]} & 
\colhead{$\Delta V_{1.1}$} &
\colhead{$\Delta V_{1.2}$} &
\colhead{$\Delta V_{1.4}$} &
\colhead{(B-V)$_{0,g}$} &
\colhead{$S_{2.5}$} &
\colhead{$S_{2.0}$} &
\colhead{$(B-V)_{0,-1}$} }
\startdata
  NGC 104& -0.70 & -0.59&  1.00 &  1.47 &  2.09&  0.95&  --- &  4.82 &  1.29 \nl
 NGC 288& -1.07 & -0.85&  1.68 &  2.05 &  2.60&  0.83&  4.71&  5.59 &  1.12 \nl 
 NGC 362& -1.15 & -0.99&  1.80 &  2.11 &  2.58&  0.81&  4.50&  5.59 &  1.07 \nl 
 NGC1261& -1.09 & -0.89&  1.87 &  2.23 &  2.72&  0.83&  5.27&  6.54 &  1.07 \nl 
 NGC1466& -1.64 & -1.44&  2.23 &  2.52 &  3.00&  0.73&  5.37&  6.54 &  0.93 \nl 
 NGC1841& -2.11 & -1.91&  2.75 &  3.00 &  --- &  0.63&  6.53&  7.90 &  0.80 \nl 
 NGC1851& -1.08 & -0.88&  1.53 &  1.90 &  2.47&  0.86&  4.52&  5.41 &  1.15 \nl 
 NGC1904& -1.37 & -1.22&  2.13 &  2.47 &  2.94&  0.78&  5.86&  7.01 &  1.00 \nl 
 NGC2419& -1.97 & -1.77&  2.54 &  2.82 &  --- &  0.72&  6.78&  8.37 &  0.88 \nl 
 NGC2808& -1.15 & -0.95&  1.70 &  2.08 &  2.60&  0.85&  4.92&  6.06 &  1.10 \nl 
 NGC3201& -1.23 & -1.03&  1.93 &  2.30 &  2.89&  0.79&  5.34&  6.14 &  1.05 \nl 
 NGC4147& -1.58 & -1.38&  2.17 &  ---  &  --- &  0.76&  --- &  6.63 &  0.99 \nl 
 NGC4372& -1.94 & -1.74&  2.72 &  3.01 &  3.46&  0.68&  7.05&  8.49 &  0.83 \nl 
 NGC4590& -1.99 & -1.81&  2.52 &  2.85 &  --- &  0.71&  6.54&  7.65 &  0.88 \nl 
 NGC4833& -1.58 & -1.27&  2.05 &  2.48 &  3.06&  0.81&  6.38&  7.26 &  1.01 \nl 
 NGC5053& -2.51 & -2.31&  2.67 &  ---  &  --- &  0.67&  6.46&  7.31 &  0.84 \nl 
 NGC5272& -1.34 & -1.16&  2.12 &  2.42 &  2.85&  0.78&  5.58&  6.98 &  0.99 \nl 
 NGC5286& -1.57 & -1.37&  2.37 &  2.68 &  3.11&  0.72&  5.98&  6.99 &  0.93 \nl 
 NGC5466& -2.14 & -1.94&  2.51 &  2.78 &  --- &  0.74&  7.06&  9.44 &  0.86 \nl 
 NGC5694& -1.72 & -1.52&  ---  &  ---  &  --- &  0.70&  7.46&  8.29 &  0.87 \nl 
 NGC5824& -1.64 & -1.44&  2.46 &  2.74 &  3.18&  0.70&  6.11&  7.39 &  0.89 \nl 
 NGC5897& -1.59 & -1.44&  2.33 &  2.62 &  3.06&  0.75&  6.07&  7.55 &  0.93 \nl 
 NGC5904& -1.11 & -0.90&  1.98 &  2.30 &  2.70&  0.81&  5.20&  6.66 &  1.04 \nl 
 NGC5927& -0.46 & -0.37&  0.26 &  0.79 &  1.53&  1.06&  --- &  3.96 &  1.52 \nl 
 NGC6093& -1.41 & -1.21&  2.27 &  2.51 &  2.88&  0.70&  5.02&  6.37 &  0.92 \nl 
 NGC6121& -1.19 & -0.94&  1.42 &  1.77 &  2.28&  0.90&  4.06&  5.18 &  1.18 \nl 
 NGC6171& -0.87 & -0.70&  1.26 &  1.64 &  2.16&  0.92&  3.76&  4.92 &  1.24 \nl 
 NGC6205& -1.39 & -1.18&  2.15 &  2.51 &  2.98&  0.76&  5.78&  6.73 &  0.98 \nl 
 NGC6218& -1.37 & -1.17&  1.75 &  2.10 &  2.60&  0.85&  4.95&  6.22 &  1.08 \nl 
 NGC6229& -1.30 & -1.10&  1.82 &  2.13 &  2.53&  0.85&  4.75&  6.63 &  1.07 \nl 
 NGC6254& -1.41 & -1.25&  2.05 &  2.38 &  2.94&  0.74&  5.00&  5.74 &  1.00 \nl 
 NGC6266& -1.07 & -0.87&  2.09 &  2.44 &  2.93&  0.77&  5.55&  6.48 &  1.02 \nl 
 NGC6333& -1.56 & -1.36&  2.44 &  2.71 &  --- &  0.79&  --- &  ---  &  0.92 \nl 
 NGC6341& -2.16 & -1.95&  2.65 &  2.91 &  --- &  0.71&  7.39&  9.74 &  0.83 \nl 
 NGC6352& -0.64 & -0.50&  0.36 &  0.85 &  1.58&  1.04&  --- &  3.99 &  1.49 \nl 
 NGC6366& -0.87 & -0.70&  0.39 &  0.90 &  1.85&  1.03&  --- &  4.79 &  1.38 \nl 
 NGC6397& -1.82 & -1.65&  2.51 &  2.84 &  --- &  0.71&  6.42&  7.51 &  0.89 \nl 
 NGC6440& -0.49 & -0.40&  ---  &  ---  &  --- &  --- &  --- &  ---  &  ---  \nl
 NGC6528& -0.38 & -0.31&  ---  &  ---  &  --- &  --- &  --- &  ---  &  ---  \nl
 NGC6535& -1.53 & -1.33&  1.80 &  2.37 &  --- &  0.86&  --- &  7.31 &  1.07 \nl 
 NGC6553& -0.44 & -0.36&  ---  &  ---  &  --- &  --- &  --- &  ---  &  ---  \nl
 NGC6584& -1.30 & -1.10&  2.03 &  2.35 &  2.84&  0.77&  5.17&  6.24 &  1.01 \nl 
 NGC6637& -0.68 & -0.55&  0.80 &  1.29 &  1.97&  0.98&  3.89&  4.65 &  1.35 \nl 
 NGC6652& -0.87 & -0.70&  ---  &  ---  &  --- &  --- &  --- &  ---  &  ---  \nl
 NGC6681& -1.27 & -1.07&  1.91 &  2.26 &  2.77&  0.80&  5.20&  6.27 &  1.05 \nl 
 NGC6712& -0.88 & -0.71&  1.47 &  1.80 &  2.33&  0.85&  4.00&  4.81 &  1.19 \nl 
 NGC6717& -1.10 & -0.90&  1.62 &  1.95 &  2.45&  0.83&  4.19&  5.14 &  1.13 \nl 
 NGC6752& -1.42 & -1.21&  2.10 &  2.40 &  2.80&  0.80&  5.67&  7.25 &  1.00 \nl 
 NGC6809& -1.61 & -1.41&  ---  &  ---  &  --- &  --- &  --- &  ---  &  ---  \nl
 NGC6838& -0.70 & -0.49&  0.69 &  1.17 &  1.87&  1.00&  3.91&  4.44 &  1.39 \nl 
 NGC6934& -1.30 & -1.10&  2.28 &  2.57 &  3.03&  0.73&  5.69&  6.94 &  0.95 \nl 
 NGC6981& -1.30 & -1.10&  1.90 &  2.23 &  --- &  0.81&  5.05&  6.26 &  1.04 \nl 
 NGC7006& -1.35 & -1.15&  2.26 &  ---  &  --- &  0.76&  6.36&  7.10 &  0.98 \nl 
 NGC7078& -2.12 & -1.91&  2.59 &  2.92 &  --- &  0.69&  6.53&  7.62 &  0.86 \nl 
 NGC7099& -1.91 & -1.71&  2.49 &  2.77 &  3.17&  0.73&  6.63&  8.08 &  0.89 \nl 
 NGC7492& -1.27 & -1.07&  1.86 &  2.12 &  2.53&  0.76&  4.00&  5.05 &  1.04 \nl 
 IC4499 & -1.26 & -1.06&  2.26 &  2.63 &  --- &  0.74&  5.95&  6.75 &  0.97 \nl 
 Rup 106& -1.70 & -1.50&  2.19 &  2.56 &  3.19&  0.73&  5.55&  6.20 &  0.96 \nl 
 Arp   2& -1.64 & -1.44&  2.33 &  2.65 &  --- &  0.73&  5.97&  7.17 &  0.92 \nl 
 Ter   7& -0.64 & -0.52&  1.06 &  1.48 &  --- &  0.94&  --- &  ---  &  1.30 \nl 
 Ter   8& -1.60 & -1.40&  ---  &  ---  &  --- &  ---&  --- &  ---  & --- \nl 
\enddata
\end{deluxetable}

Over the past 15 years many calibrations of these parameters in terms
of metallicity have been proposed (see Table 5 in Ferraro, Fusi Pecci
\& Buonanno 1992 and, recently, SL97, Carretta \& Bragaglia 1998, and
Table 7 by Borissova et al 1999). We present here revised calibrations
making use of the wider and more homogeneous data-set now available.
The six parameters defined above (namely, $\Delta V_{1.1}$, $\Delta
V_{1.2}$, $\Delta V_{1.4}$, $(B-V)_{0,g}$, $S_{2.0}$, $S_{2.5}$) have
been measured for all the GGCs listed in our catalog and having
suitable data in the $(V,~B-V)$ plane. No attempt has been performed
to extrapolate the mean ridge line beyond the sufficiently populated
regions of the available CMDs.  The values measured are listed in
Table 3.  The main source of uncertainty in these measures
is the propagation of the the error in the determination of the ZAHB
level. Thus, the uncertainty on the $V_{\rm ZAHB}$ ($\sim 0.10$)
typically produces a comparable uncertainty ($\delta \sim 0.1$
mag) in measuring $\Delta V$-parameters, an error $\delta \sim 0.03$
mag in the determination of the colour $(B-V)_{0,g}$, and a
significantly larger error ($\delta \sim 0.2-0.3$) in deriving the $S$
parameters.

In Figure 6--11 a,b, the RGB-observables are plotted versus the metal
abundances $\feh_{\rm CG97}$ (panel (a)) and the global metallicities
$\mh$ (panel (b)).  The solid line overplotted in each Figure the best
fits to the data given in analytic form in Table 4. In deriving these
relations, the 20 GGCs with direct spectroscopic measures of \feh\ and
direct measures of the $\alphafe$ abundance (excepting NGC 7099) have
been considered as {\it primary calibrators} and have been assumed
with higher weights in determining the best fit relations. These
primary calibrators are plotted as filled circles.  Two clusters
(NGC~5053 and NGC~5694) which have the largest deviations in Figure 10
(plotted as dotted circles) have been excluded during the fitting
procedure. In both of the excluded clusters the observed samples the
upper RGB is so poorly populated that the location of the branch at
that level is quite uncertain.  NGC~5053 has been excluded, for the
same reason, in deriving the calibration of the parameter
$S_{2.0}$ (see Figure 11).  The relationships of all the RGB
parameters defined above in terms of the spectroscopic and {\it
global} metallicity scales are reported in Table 4.  Note that the
quoted relations can be safely used only in the metallicity range
covered by the adopted sample (i.e. roughly $-2.5<\feh_{\rm
CG97}<-0.5$ and $-2.3<\mh<-0.4$). This range should be considered as a
first guess and the reader is requested to refer to each figure
(Figure 6--11) to check the exact range of metallicity within which
each relation has been derived.

\begin{deluxetable}{ccc}
\hoffset=-5truemm
\tablewidth{17.5truecm}
\label{lm}
\tablecaption{RGB parameters and their calibration in terms of different
metallicity scales.}
\tablehead{
\colhead{Relations in terms of $\feh_{\rm CG97}$} &
\colhead{} &
\colhead{} }
\startdata
 $\feh_{\rm CG97} = -0.315 \Delta V_{1.1}^2+0.347 \Delta V_{1.1} -0.768$ 
& n=54 $\sigma=0.15$& (4.1)\nl
 $\feh_{\rm CG97} = -0.359 \Delta V_{1.2}^2+0.708 \Delta V_{1.2} -1.023$ 
& n=51 $\sigma=0.14$&(4.2)\nl
 $\feh_{\rm CG97} = -0.252 \Delta V_{1.4}^2+0.548 \Delta V_{1.4} -0.864$ 
& n=38 $\sigma=0.12$& (4.3)\nl
 $\feh_{\rm CG97} = -9.47 (B-V)_{0,g}^2 + 20.127 (B-V)_{0,g} -11.36$ 
& n=55 $\sigma=0.20$&(4.4)\nl
 $\feh_{\rm CG97} = -0.37 S_{2.5} +0.59$  & n=45 $\sigma=0.18$ &(4.5)\nl
 $\feh_{\rm CG97} = -0.28 S_{2.0} +0.67$  & n=52 $\sigma=0.18$ &(4.6)\nl
 $(B-V)_{0,g} = 0.005 \feh_{\rm CG97}^3+0.118 \feh_{\rm CG97}^2+0.489 
\feh_{\rm CG97}+1.243$& n=55 $\sigma=0.04$ & (4.7)\nl
    & & \nl
\hline
   Relations in terms of $\mh$ &  \nl
\hline\nl
 $\mh = -0.337 \Delta V_{1.1}^2+0.434 \Delta V_{1.1} -0.656$ & 
 n=54 $\sigma=0.14$ &(4.8)\nl
 $\mh = -0.382 \Delta V_{1.2}^2+0.820 \Delta V_{1.2} -0.960$ &
 n=51 $\sigma=0.13$ &(4.9)\nl
 $\mh = -0.280 \Delta V_{1.4}^2+0.717 \Delta V_{1.4} -0.918$ &
 n=38 $\sigma=0.12$ &(4.10)\nl
 $\mh = -10.513 (B-V)_{0,g}^2 + 21.813 (B-V)_{0,g} -11.835$ & 
 n=55 $\sigma=0.19$ &(4.11)\nl
 $\mh = -0.36 S_{2.5} +0.78$  & n=45 $\sigma=0.18$ &(4.12)\nl
 $\mh = -0.29 S_{2.0} +0.53$  & n=52 $\sigma=0.18$ &(4.13)\nl
 $(B-V)_{0,g} = 0.04 \mh^3+0.275 \mh^2+0.67 \mh +1.252$&
n=55 $\sigma=0.04$ & (4.14)\nl
    &  &\nl
\hline
   The parameter $(B-V)_{0,-1}$ vs metallicity &  \nl
\hline\nl
$ (B-V)_{0,-1} = 0.055\feh^3_{CG97}+0.448\feh^2_{CG97}+
 +1.255\feh_{\rm CG97}+2.023 $ & n=55 $\sigma=0.05$ &(4.15)\nl
 $(B-V)_{0,-1} = 0.115\mh^3+0.695\mh^2+ 1.496\mh+1.983$ &
 n=55 $\sigma=0.05$&(4.16)\nl
\enddata
\end{deluxetable}

Similar relations for $\Delta V_{1.1}$, $\Delta V_{1.2}$ and
$(B-V)_{0,g}$ have been recently obtained by Carretta \& Bragaglia
(1998), who used the values listed by SL97 and SN94. For sake of
comparison, their relations have been plotted as dashed lines in the
corresponding Figures. The differences can easily be understood as due
to the combination of two main factors:
 
\begin{enumerate}

\item The different assumptions on the HB level; there are differences up to  
0.15 mag between the HB levels adopted here and those listed in SL97 and SN94.
These differences directly affect the measure of $\Delta V$ parameters, and 
this explains most of the offset between the solid and the dashed lines in 
Figure 5a,6a.

\item The small sample (only 9 GGCs) considered by Carretta \& Bragaglia (1998);
this mainly affects the fit at the extremes. In fact, the dashed lines 
significantly deviate from the solid line at the extremes of the metallicity 
scale (see Figure 6a,7a,9a).

\end{enumerate}

\subsection {The SRM in the $(V,~B-V)$ plane}

The equations reported in Table 4 represent a system which can be used 
to {\it simultaneously} derive very useful estimates  of metal abundance 
($\feh_{\rm CG97}$ and $\mh$) and reddening from the morphology and 
location  of the RGB (the so-called SRM Method,  Sarajedini 1994b). 
By using the system of equations in Table 4 one can choose
the most appropriate observables (measurable in the CMD depending on
the actual extension of the observed RGB) and then proceed as follows:

\begin{enumerate}
\item Since parameters $S_{2.0}$ and $S_{2.5}$ are independent 
of cluster reddening, from eq.s  4.5 and 4.6 it is possible to obtain 
a first guess of the cluster metallicity $\feh_i$, and similarly $\mh_i$
using  eq.s 4.12 and 4.13.

\item Introducing then $\feh_i$ in eq.  4.7 it is possible to 
derive a first value for the expected $(B-V)_{0,g}$ and, in turn, a 
first estimate for the reddening from $E(B-V)=(B-V)_g-(B-V)_{0,g}$

\item Using this first estimate of the reddening it is then
possible to derive 
$\Delta V_{1.1}$, 
$\Delta V_{1.2}$, 
$\Delta V_{1.4}$, and from eq.s  4.1, 4.2 and 4.3 a new determination of 
the metallicity.

\item By iterating the procedure one can quickly achieve
convergence,  yielding values for reddening and the metallicities 
generally accurate at about $\delta \feh<0.1$ and $\delta E(B-V)<0.02$.

\end{enumerate}

\begin{figure}
\epsffile{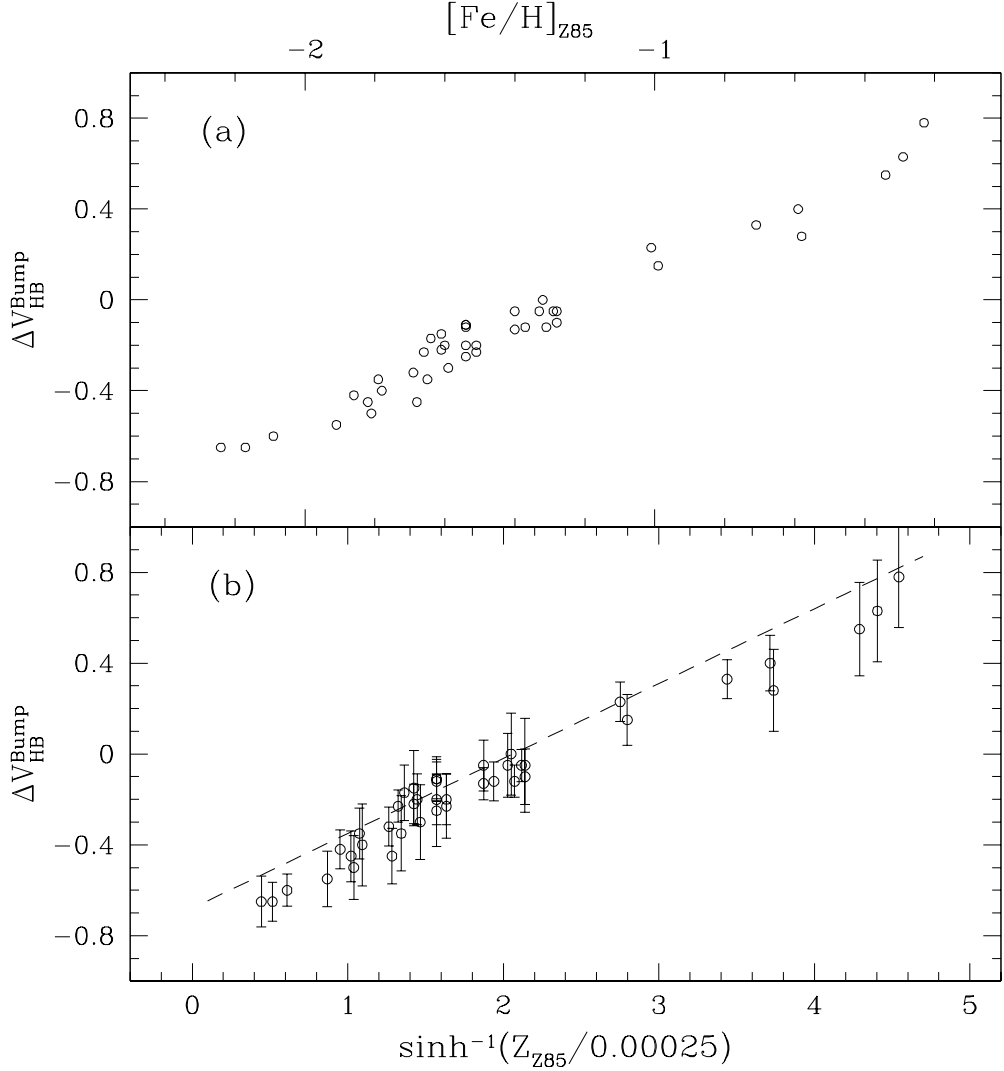}
\caption{
\protect\label{bump1}
The 
$\Delta V_{\rm HB}^{Bump}$ parameter
as function of the metallicity ($\feh_{\rm Z85}$) 
and the parameter
$s$ 
{\it (panel (a)} and  
{\it (b)}, respectively).
The dashed line is the relation obtained by F90.
}
\end{figure}

\begin{figure}
\epsffile{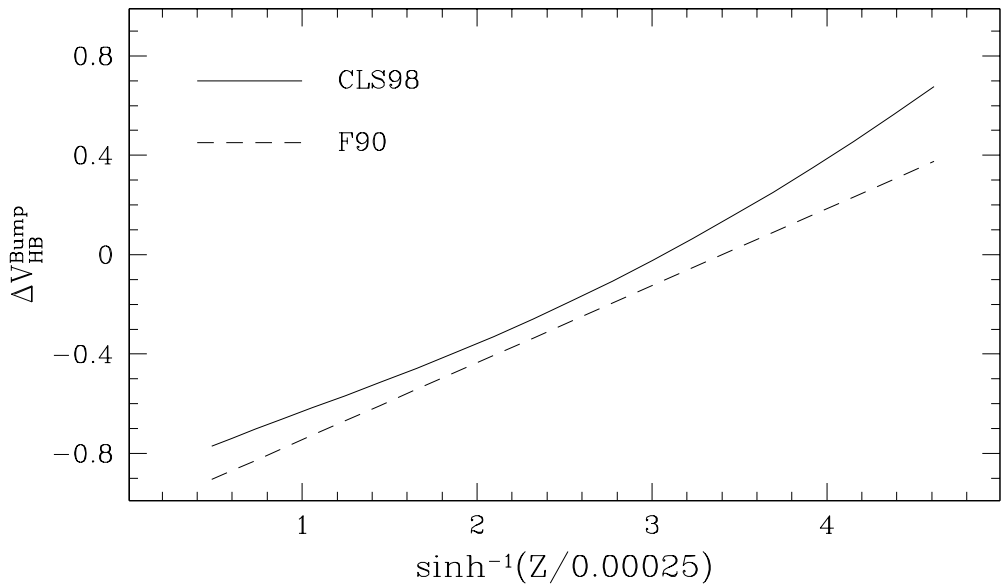}
\vspace{-5cm}
\caption{
\protect\label{bump1}
{\it New} and {\it old} theoretical $\Delta V_{\rm HB}^{Bump}$ 
as function of $s$:
present models SLC98 (solid line) and 
the relation adopted by F90 (dashed line).
}
\end{figure}

\begin{figure}
\epsffile{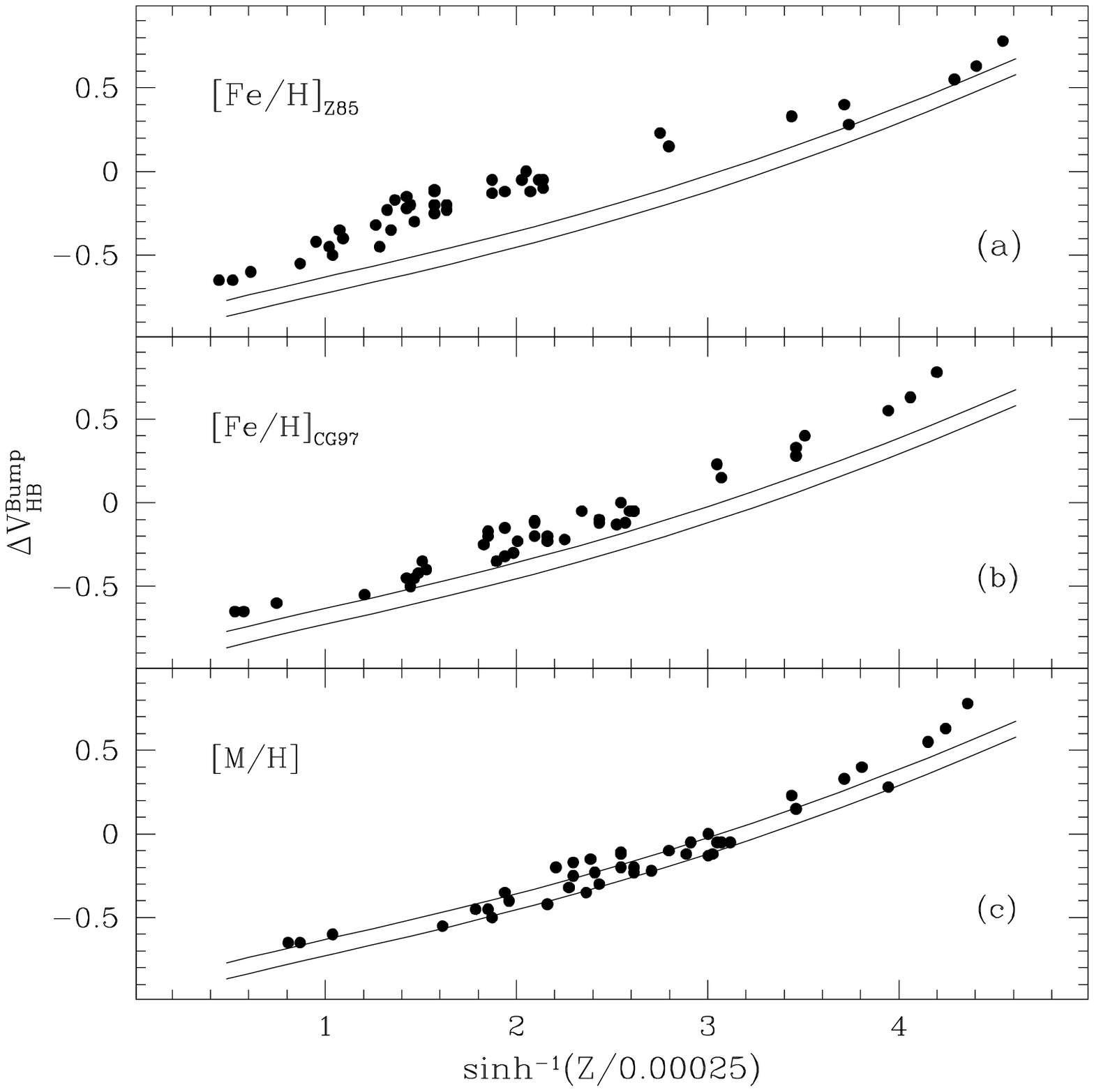}
\vspace{-1cm}
\caption{
\protect\label{bump1}
The same as Figure 12 for the three different 
metallicity scales:
Zinn 1985 (upper panel), 
Carretta \& Gratton 1997 (central panel), {\it global} metallicity
$\mh$ (lower panel). 
The two solid line represent the 
theoretical predictions for two different ages, namely 12 
(lower line)
and 16 Gyr (upper line).
}
\end{figure}

\subsection {The RGB-bump: the new data-base}

One of the most intriguing features along the RGB is the so-called
RGB-bump whose existence was predicted since the early theoretical
models (Thomas 1967, Iben 1968)
 but was first observed years later (King, Da Costa \&
Demarque, 1985) the observed samples were not populous enough to allow
a firm detection. In fact a very large sample of stars, with more than
1000 in the upper 4 mag. of the RGB, is necessary to safely
distinguish this feature from statistical fluctuations.  This problem
is more severe for metal poor clusters, due to the dependance of the
luminosity of the bump on metallicity: the luminosity increases with
decreasing metallicity.  So, in metal poor GGCs the bump is shifted
toward the RGB tip, in a region which is intrinsically poorly
populated, and where the detection is difficult even when large
samples are available.

The first systematic study of the location of the RGB-Bump in GGCs has
been presented by Fusi Pecci et al. (1990, hereafter F90), who
reported the identification of the bump in a sample of 11 GGCs.  They
presented a detailed comparison with theoretical models (Rood \&
Crocker 1989, Rood unpublished) based on old input physics.  This
first comparison showed that while the dependance of bump luminosity
on metallicity was nicely reproduced by theoretical models, there
was a substantial disagreement in the zero-point, the theoretical
relation being about 0.4 mag brighter than the observations.  Alongi
et al. (1991) interpreted this disagreement as an evidence of the
limit of the standard models in describing the correct location of the
RGB-bump. Thus, in order to reconcile observations and theory, they
claimed the occurrence of an additional mixing process below the
bottom of the convective envelope of an RGB star (i.e. {\it
under-shooting}).  However, Straniero, Chieffi \& Salaris (1992) and
Ferraro (1992), independently pointed out that a proper inclusion of
the $\alpha-$element enhancement in the computation of the global
metallicity of the parent cluster could reduce the discrepancy. In a
recent review of the problem, Cassisi \& Salaris (1997) essentially
re-obtained the same result.

Since the early work presented in F90, the RGB-Bump has been
identified in a growing number of GGCs (see for example Brocato et
al. 1996).  As pointed out by Rood \& Crocker (1985), the best tool to
identify the RGB-bump is the Luminosity Function (LF), and both the
integrated and the differential LFs are useful (Ferraro
1992). Following the prescriptions of F90 we independently identified
the RGB-bump in 47 GGCs in our catalog.  The bump magnitudes so
measured are listed in column 5 of Table 5.  This represents the
largest GGCs sample listing the RGB-bump locations available so far.

To allow comparisons with both previous studies and theoretical
models, following F90, we have measured the parameter: $\Delta
V_{\rm HB}^{\rm Bump}=V_{\rm Bump}-V_{\rm HB}$, which has the
advantage of being actually independent of the photometric zero-point
of the cluster data, the reddening, and the distance modulus.
Moreover we have also adopted the parameter $s$ defined as: $s=
{\rm sinh}^{-1}(Z/0.00025)$, where $Z=10^{\feh-1.7}$.  This quantity is best
suited to linearly describe the dependence of $\Delta V_{\rm HB}^{\rm
Bump}$ on metallicity (F90).

As a starting point, we first adopt the same metallicity scale
as F90, namely the Zinn (1985) scale.  In Figure 12a, we reported the
$\Delta V_{\rm HB}^{\rm Bump}$ as a function of $\feh$, while in
Figure 12b the same quantity is plotted versus $s$.  As can be seen
there is a clearcut correlation.  The error bars, as expected, tend to
be systematically larger at lower metallicities, due to the
difficulty, mentioned above, in identifying the bump in those GGCs.

The best fit to the data obtained by F90 is also overplotted to the
new values in the same figure. As can be seen there is a systematic
shift, the new values being slightly lower the the old ones by about 
0.05-0.1 mag.  Such a difference is mainly due to the new procedure
used to measure the ZAHB level. In fact in F90 the lower edges of the
observed HB distributions were assumed to be coincident with the ZAHB
levels (see the discussion presented in section 4).

On the theoretical side, the latest models which include improvements
in the input physics imply a reduction of the predicted luminosity
level of the RGB bump (see e.g. table 4 in SCL97). In addition to
that, larger core masses at the He flash are now obtained, so that the
predicted HB luminosity is larger than those found in the old
computations.  On the base of the RGB and HB models described in
Section 4, we have derived the following relation for the RGB-bump
location:

\begin{equation}
\label{eq:2}
{\rm M_V^{\rm Bump}} = 0.7502 +0.9896 \log\,t_9 +1.5797 \mh 
+0.2574 \mh^{2}\\
\end{equation}

where $t_9$ is the age in Gyr, and for the ZAHB level:

\begin{equation}
\label{eq:3}
M_V^{\rm ZAHB} = 1.0005 +0.3485 \mh +0.0458 \mh^{2}\\
\end{equation}

 In order to make easy the comparison with other published relation 
for the ZAHB level as a function of the metallicity,
it could be useful to give here also the linear best-fit
regression in the range $-0.4<[M/H]<-2.2$:

\begin{equation}
\label{eq:3}
M_V^{\rm ZAHB} = 0.23 \mh + 0.94\\
\end{equation}

In Figure 13 we show a comparison between the old theoretical values
(adoped by F90)
for the $\Delta V_{\rm HB}^{\rm Bump}$ parameters and the latest
ones. The new values are significantly larger by about 0.15--0.20 mag
(up to 0.3 mag at the largest metallicity).  Results from the new
models are compared to the present data in Figure 14.  We show the
theoretical expectations for two different ages, 12 and 16 Gyr, which
are roughly representative of the range of ages covered by the bulk of
the Galactic Globular Clusters. The ``old'' (Z85) and the ``new''
(CG97) metallicity scale are shown, in panel a) and b) respectively.
Finally in panel c) we adopt the {\it global} metallicity ($\mh$).
Note that only in this third case a good agreement between the theory
and the observations is obtained. 
The previous
discrepancy of about 0.4 mag between theory and observation has been
completely removed. The major changes are:

\begin{itemize}

\item The updated input physics in the evolutionary models 
which yields: RGB bump less luminous (by $\sim 0.1-0.15$ mag)
because of the increased opacity, and HB level more luminous 
(by $\sim 0.05$ mag) because of increased core mass.

\item The new spectroscopic abundances ($\alphafe$ and $\feh$) 
which contribute $\sim 0.2$ mag.

\item The new definition of the HB level, 
which contributes $\sim 0.05-0.1$ mag.

\end{itemize}

The best fit relations obtained in terms of the different
metallicity scales  are listed in Table 6. For each metallicity scale
the behaviour
of the $\Delta V_{\rm HB}^{\rm Bump}$ parameter  has been computed 
both in terms of the usual metallicity parameter ($\feh, \mh$ etc)
and the parameter $s$ defined by F90.

\section{ THE ASYMPTOTIC GIANT BRANCH}

According to the evolutionary models (Castellani, Chieffi \& Pulone,
1991), after the exhaustion of the central He, the He-burning rapidly
move from the center toward the maximum mass coordinate attained by
the convective core during the HB phase.  Thus, the beginning of the
AGB is characterized by a rapid increase of the luminosity.  When the
shell He burning stabilizes, a slowing down in the
evolutionary rate is expected. Then, from an observational point of
view, the transition between the central and the shell He-burning
should be marked by a clear gap (where few stars should be found),
while a well defined clump of stars should indicate the base of the
AGB.

It has been recognized (Castellani, Chieffi \& Pulone 1991, Pulone
1992, Bono et al. 1995) that the luminosity level of the AGB-clump is
almost independent of the chemical composition of the cluster stars
(both $Z$ and $Y$), so that this (quite bright) feature could be a very
promising ``standard candle.'' However, we note that the
theoretical calibration of the AGB clump location is affected by the
uncertainties in the actual extension of the convective core of an He
burning low mass stars. On the other hand, as pointed out by Caputo et
al. (1989) one might use the observed differences between the HB
luminosity level and that of the AGB-clump (i.e. $\Delta V_{\rm AGB}^{\rm
HB} =V^{\rm AGB}_{clump}-V_{\rm HB}$) to constrain the convection theory
(instability criterion, semiconvection, overshooting and the like; see
Dorman \& Rood 1993).

Unfortunately, the identification of such a clump is not easy since
the AGB phase itself is very short ($\sim 10^7$ yr) and, in turn,
always poorly populated (a GGC with total luminosity $L_T=10^5\lsun$
contains $\sim 20$ AGB stars; see Renzini \& Fusi Pecci 1988).  There
are a few identifications of the AGB-clump in the literature: Ferraro
(1992) reported a preliminary identification of this feature in 3 GGCs
(M5, NGC 1261, NGC 2808), and Montegriffo et al (1995) showed that it
is clearly visible in 47 Tuc. Other examples could be found in
published CMD's, but the AGB-clump detection has neither been noted nor
discussed.

To initiate a systematic study of the properties of the AGB-clump,
we have independently identified such a feature in 9 GGCs whose CMDs
show a significant clump of stars in the AGB region.
In Table 5 
(column 6), the apparent $V$ magnitudes
of the approximate centroid 
of the AGB-clump stars 
distribution
are listed for each of these 9 clusters.

In order to study the behaviour of the AGB-clump from a theoretical
point of view, a subset of CLS98 models were continued through the
onset of He shell burning up the the of the first thermal pulse.  The
main results are presented in Figure 15 where the absolute $V$
magnitude of the AGB-clump is plotted versus the ZAHB mass, for two
different metallicities ($\log Z=-3$ and $\log Z=-4$, respectively).
The corresponding $(B-V)$ colours of the AGB-clump are also reported
for each model.  Both the colour and the luminosity of the AGB-clump
depend significantly on the stellar mass (see also Fig 2 by
Castellani, Chieffi \& Pulone 1989).

\clearpage
\hoffset = -1mm
\begin{deluxetable}{cccccc}
\tablewidth{15truecm}
\tablecaption{RGB and AGB Bump parameters.}
\tablehead{
\colhead{ $Name$} & 
\colhead{[Fe/H]$_{CG97}$} & 
\colhead{[M/H]} & 
\colhead{$V_{ZAHB}$} &
\colhead{$V_{RGB}^{Bump}$} &
\colhead{$V_{AGB}^{Bump}$}  }
\startdata
 NGC 104& -0.70 & -0.59& 14.22$\pm$0.07 & 14.55$\pm$0.05&  13.15$\pm$0.07 \nl 
 NGC 288& -1.07 & -0.85& 15.50$\pm$0.10 & 15.45$\pm$0.05&  --- \nl 
 NGC 362& -1.15 & -0.99& 15.50$\pm$0.07 & 15.40$\pm$0.10&  --- \nl 
 NGC1261& -1.09 & -0.89& 16.72$\pm$0.05 & 16.60$\pm$0.05&  15.65$\pm$0.05 \nl 
 NGC1466& -1.64 & -1.44& 19.30$\pm$0.07 &      ---    &  --- \nl 
 NGC1841& -2.11 & -1.91& 19.42$\pm$0.10 &      ---     &  18.55$\pm$0.10 \nl 
 NGC1851& -1.08 & -0.88& 16.20$\pm$0.05 & 16.15$\pm$0.05&  --- \nl 
 NGC1904& -1.37 & -1.22& 16.27$\pm$0.07 & 15.95$\pm$0.05&  --- \nl 
 NGC2419& -1.97 & -1.77& 20.50$\pm$0.10 &      ---     &  --- \nl 
 NGC2808& -1.15 & -0.95& 16.27$\pm$0.07 & 16.15$\pm$0.05&  15.20$\pm$0.07 \nl 
 NGC3201& -1.23 & -1.03& 14.77$\pm$0.07 & 14.55$\pm$0.05&  --- \nl 
 NGC4147& -1.58 & -1.38& 16.95$\pm$0.10 &      ---     &  --- \nl 
 NGC4372& -1.94 & -1.74& 15.90$\pm$0.15 &      ---     &  --- \nl 
 NGC4590& -1.99 & -1.81& 15.75$\pm$0.05 & 15.15$\pm$0.05&  --- \nl 
 NGC4833& -1.58 & -1.27& 15.77$\pm$0.07 & 15.35$\pm$0.05&  --- \nl 
 NGC5053& -2.51 & -2.31& 16.70$\pm$0.07 &      ---     &  --- \nl 
 NGC5272& -1.34 & -1.16& 15.68$\pm$0.05 & 15.45$\pm$0.05&  14.80$\pm$0.05 \nl 
 NGC5286& -1.57 & -1.37& 16.60$\pm$0.10 & 16.25$\pm$0.05&  15.57$\pm$0.10 \nl 
 NGC5466& -2.14 & -1.94& 16.62$\pm$0.10 &      ---     &  --- \nl 
 NGC5694& -1.72 & -1.52& 18.70$\pm$0.10 & 18.15$\pm$0.07&  --- \nl 
 NGC5824& -1.64 & -1.44& 18.52$\pm$0.07 &      ---     &  --- \nl 
 NGC5897& -1.59 & -1.44& 16.45$\pm$0.07 & 16.00$\pm$0.10&  --- \nl 
 NGC5904& -1.11 & -0.90& 15.13$\pm$0.05 & 15.00$\pm$0.05&  14.15$\pm$0.05 \nl 
 NGC5927& -0.46 & -0.37& 16.72$\pm$0.10 &      ---     &  --- \nl 
 NGC6093& -1.41 & -1.21& 16.12$\pm$0.07 & 15.95$\pm$0.10&  --- \nl 
 NGC6121& -1.19 & -0.94& 13.45$\pm$0.10 & 13.40$\pm$0.10&  --- \nl 
 NGC6171& -0.87 & -0.70& 15.70$\pm$0.10 & 15.85$\pm$0.05&  --- \nl 
 NGC6205& -1.39 & -1.18& 15.10$\pm$0.15 & 14.75$\pm$0.07&  --- \nl 
 NGC6218& -1.37 & -1.17& 14.75$\pm$0.15 & 14.60$\pm$0.07&  --- \nl 
 NGC6229& -1.30 & -1.10& 18.11$\pm$0.05 & 18.00$\pm$0.07&  17.15$\pm$0.05 \nl 
 NGC6254& -1.41 & -1.25& 14.85$\pm$0.10 & 14.65$\pm$0.05&  --- \nl 
 NGC6266& -1.07 & -0.87& 16.40$\pm$0.20 & 16.35$\pm$0.05&  --- \nl 
 NGC6333& -1.56 & -1.36& 16.35$\pm$0.15 & 15.95$\pm$0.10&  --- \nl 
 NGC6341& -2.16 & -1.95& 15.30$\pm$0.10 & 14.65$\pm$0.05&  --- \nl 
 NGC6352& -0.64 & -0.50& 15.30$\pm$0.10 &      ---     &  --- \nl 
 NGC6366& -0.87 & -0.70& 15.80$\pm$0.10 &      ---     &  14.75$\pm$0.10 \nl 
 NGC6397& -1.82 & -1.65& 13.00$\pm$0.10 &      ---     &  --- \nl 
 NGC6440& -0.49 & -0.40& 18.70$\pm$0.20 & 19.25$\pm$0.05&  --- \nl 
 NGC6528& -0.38 & -0.31& 17.17$\pm$0.20 & 17.95$\pm$0.10&  --- \nl 
 NGC6535& -1.53 & -1.33& 15.90$\pm$0.15 &      ---     &  --- \nl 
 NGC6553& -0.44 & -0.36& 16.92$\pm$0.20 & 17.55$\pm$0.10&  --- \nl 
 NGC6584& -1.30 & -1.10& 16.60$\pm$0.05 & 16.40$\pm$0.10&  --- \nl 
 NGC6637& -0.68 & -0.55& 15.95$\pm$0.10 & 16.35$\pm$0.07&  --- \nl 
 NGC6652& -0.87 & -0.70& 16.07$\pm$0.10 &      ---     &  --- \nl 
 NGC6681& -1.27 & -1.07& 15.85$\pm$0.10 & 15.65$\pm$0.05&  --- \nl 
 NGC6712& -0.88 & -0.71& 16.32$\pm$0.07 & 16.55$\pm$0.05&  --- \nl 
 NGC6717& -1.10 & -0.90& 15.75$\pm$0.15 & 15.75$\pm$0.10&  --- \nl 
 NGC6752& -1.42 & -1.21& 13.90$\pm$0.15 & 13.65$\pm$0.05&  --- \nl 
 NGC6809& -1.61 & -1.41& 14.60$\pm$0.10 & 14.15$\pm$0.05&  --- \nl 
 NGC6838& -0.70 & -0.49& 14.52$\pm$0.10 & 14.80$\pm$0.15&  --- \nl 
 NGC6934& -1.30 & -1.10& 16.97$\pm$0.07 & 16.85$\pm$0.05&  --- \nl 
 NGC6981& -1.30 & -1.10& 16.86$\pm$0.07 & 16.75$\pm$0.07&  --- \nl 
 NGC7006& -1.35 & -1.15& 18.85$\pm$0.15 & 18.55$\pm$0.07&  --- \nl 
 NGC7078& -2.12 & -1.91& 15.90$\pm$0.07 & 15.25$\pm$0.05&  --- \nl 
 NGC7099& -1.91 & -1.71& 15.30$\pm$0.10 &      ---     &  --- \nl 
 NGC7492& -1.27 & -1.07& 17.78$\pm$0.10 & 17.55$\pm$0.10&  --- \nl 
 IC4499& -1.26 & -1.06& 17.70$\pm$0.07 &      ---     &  --- \nl 
 Rup 106& -1.70 & -1.50& 17.85$\pm$0.10 &      ---     &  --- \nl 
 Arp   2& -1.64 & -1.44& 18.30$\pm$0.15 &      ---     &  --- \nl 
 Ter   7& -0.64 & -0.52& 17.87$\pm$0.10 &      ---     &  --- \nl 
 Ter   8& -1.60 & -1.40& 18.15$\pm$0.10 & 17.65$\pm$0.10&  --- \nl 
\enddata
\end{deluxetable}

 In particular higher stellar
masses tend to generate brighter and redder AGB clump. Thus, in
principle, the dependence of the AGB-clump luminosity on the mass of
the evolving star implies, in turn, 
an indirect dependence on all the other parameters
which could affect the mean mass (and its distribution) along the HB:
namely metallicity, age, mass loss efficiency and all the parameters
which directly or indirectly affect the mass loss process.  However,
it is interesting to note that for low metallicity clusters (but in
general for clusters with blue HB) the AGB clump rapidly tends to
become bluer and bluer (up to $(B-V)\sim 0.0$) and probably (due to
the spread in mass along the ZAHB) progressively less {\it clumpy}
and, for this reason, less observable. This effect is nicely shown, 
for example,
by Fig 4 
in  Rood, Whitney \& D'Cruz 1997 and Fig. 8 in Whitney \etal (1998).
Thus, operatively, whether the AGB clump is observable or
not determines to some extent the possible range luminosity which
might be observable for this feature. The models suggest that
observable AGB-Bumps are located at $M_V^{\rm AGB-Bump}=-0.3\pm 0.1$

\begin{deluxetable}{cccc}
\tablewidth{15truecm}
\label{lm}
\tablecaption{RGB-Bump parameters and their calibration in terms of the different 
metallicity scales.}
\tablehead{
\colhead{ }&
\colhead{The Zinn Scale   $\feh_{\rm Z85}$ }& 
\colhead{ }& \colhead{ }
}
\startdata
&   $\Delta V_{\rm HB}^{\rm Bump}= 0.31s_{Z85}-0.72$ & 
n=42 $\sigma=0.07$ &(6.1) \nl
& $\Delta V_{\rm HB}^{\rm Bump}= 0.67\feh_{\rm Z85}+0.827$& 
n=42 $\sigma=0.06$ &(6.2)\nl
& & &\nl
\hline
&     The Carretta-Gratton Scale   $\feh_{\rm CG97}$ &   &\nl
\hline\nl
 & $\Delta V_{\rm HB}^{\rm Bump}= 0.041s_{CG97}^2+0.172s_{CG97}-0.753$ 
& n=42 $\sigma=0.06$ &(6.3) \nl
 & $\Delta V_{\rm HB}^{\rm Bump}= 0.269\feh^2_{CG97}+1.451\feh_{\rm CG97}
+1.220$ & n=42 $\sigma=0.06$ &(6.4)\nl
 & & &\nl
\hline
&     The {\it global} Scale   $\mh$ &  &  \nl
\hline\nl
& $\Delta V_{\rm HB}^{\rm Bump}= 0.065s^2_{global}+0.025s_{global}-0.702$&
 n=42 $\sigma=0.07$ & (6.5)\nl
& $\Delta V_{\rm HB}^{\rm Bump}= 0.360\mh^2+1.602\mh+1.113$ &
n=42 $\sigma =0.07$ & (6.6)\nl
& & &\nl
\hline
 &  The Bump location in the absolute plane $M_V^{\rm Bump}$ & \nl
\hline\nl
&$M_V^{\rm Bump} = 0.29\feh^2_{CG97}+
1.736\feh_{\rm CG97}+2.23$ & n=42 $\sigma=0.06$ &(6.7) \nl
& $M_V^{\rm Bump} = 0.406\mh^2+1.95\mh+2.113$ & n=42 $\sigma=0.06$ &(6.8)\nl
\enddata
\end{deluxetable}

 \begin{figure}
\epsffile{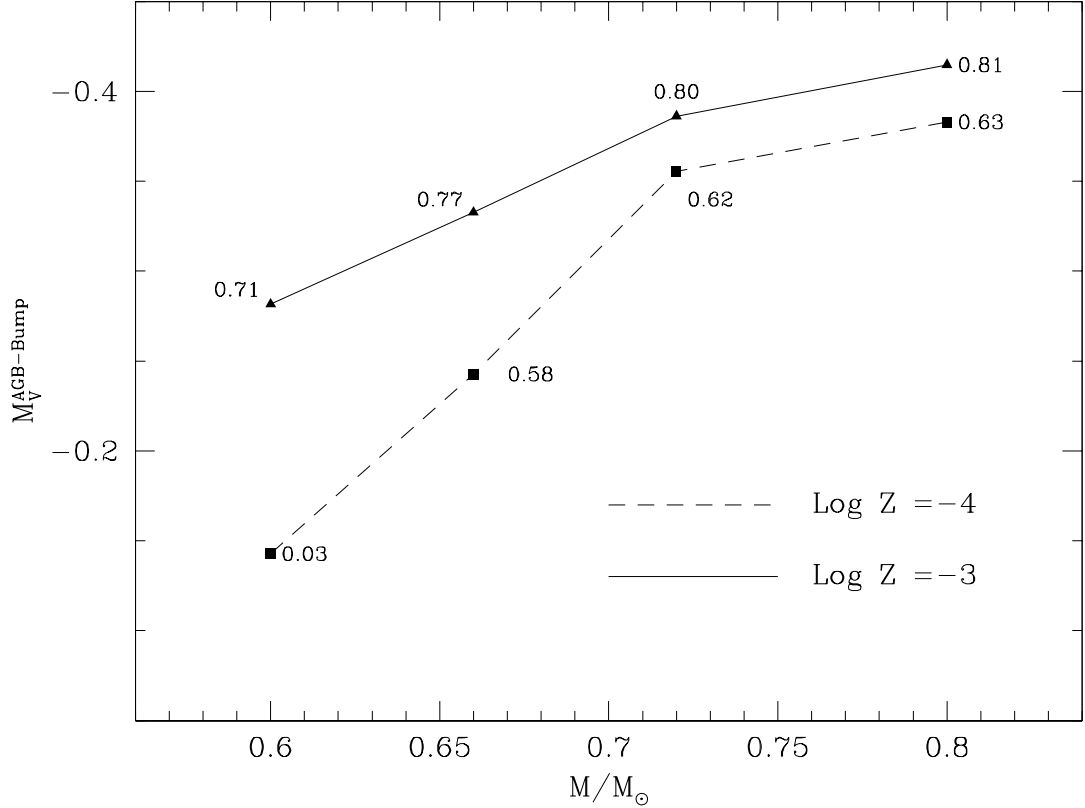}
\vspace{-2cm}
\caption{
\protect\label{agb1}
The theoretical luminosity level of the AGB clump as function of the stellar mass for
two metallicities, namely Z=0.0001 and Z=0.001
(dashed and solid line, respectively). 
The labels reported in the figure are the
B-V colours of the computed AGB clumps.
}
\end{figure}

\begin{figure}
\epsffile{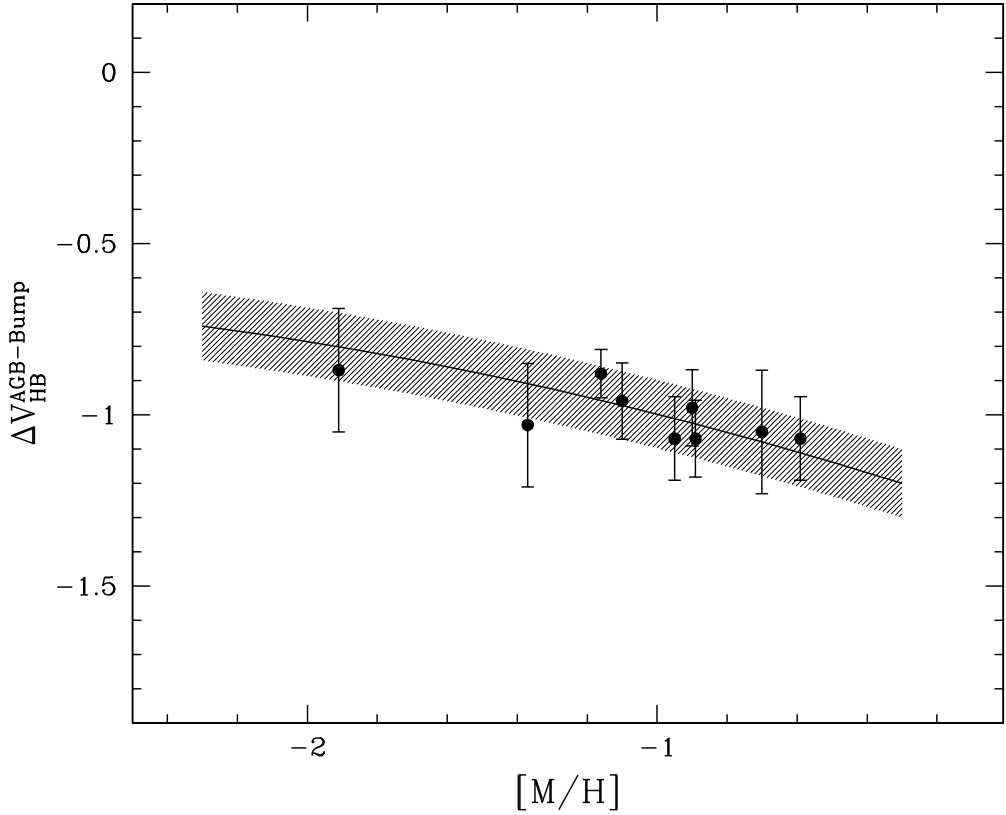}
\vspace{-2cm}
\caption{
\protect\label{agb2}
The difference between the observed ZAHB and AGB clump luminosity levels of
9 clusters in our catalog. The solid line is the theoretical
expectation.
The dashed region is 
representative of the uncertainty ($\pm 0.1$ mag) in the absolute location
of the AGB-clump (see text).
}
\end{figure}

\begin{figure}
\epsffile{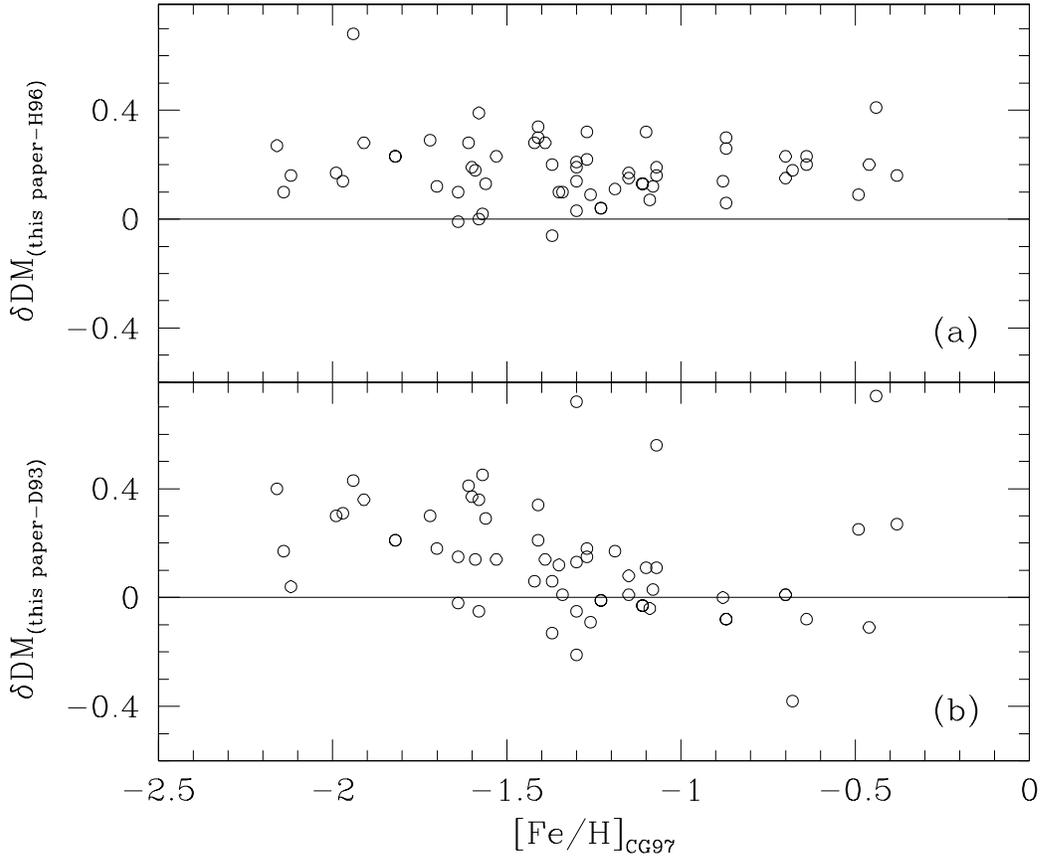}
\caption{
\protect\label{DM}
Difference between the DM obtained in this paper and the previous 
compilations by H96 and D93 ( {\it panel (a)} and {\it (b)}, respectively).
A systematic difference of $\sim 0.2$ mag is evident in the comparison with H96
data. The trend with the metallicity clearly evident in {\it panel (b)}
is due to the D93 assumption  on the $V(HB)$, (see text).
}
\end{figure}

\begin{figure}
\epsffile{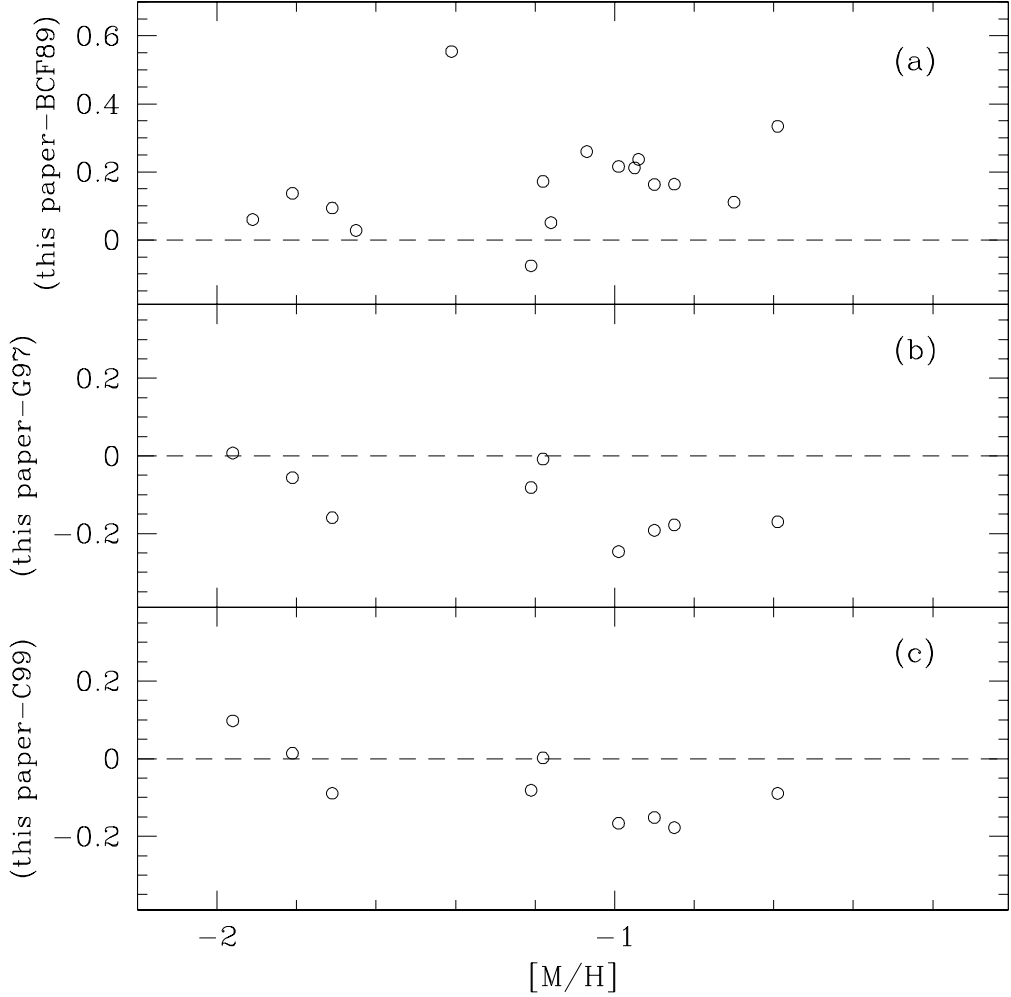}
\caption{
\protect\label{DM}
Difference between the DM obtained in this paper and those obtained from:
{\it panel (a)} BCF89;
{\it panel (b)} Hipparcos parallaxes (Gratton et al 1997);
{\it panel (c)} Revided Hipparcos parallaxes (Carretta et al 1999).
}
\end{figure}

\begin{figure*}
\epsffile{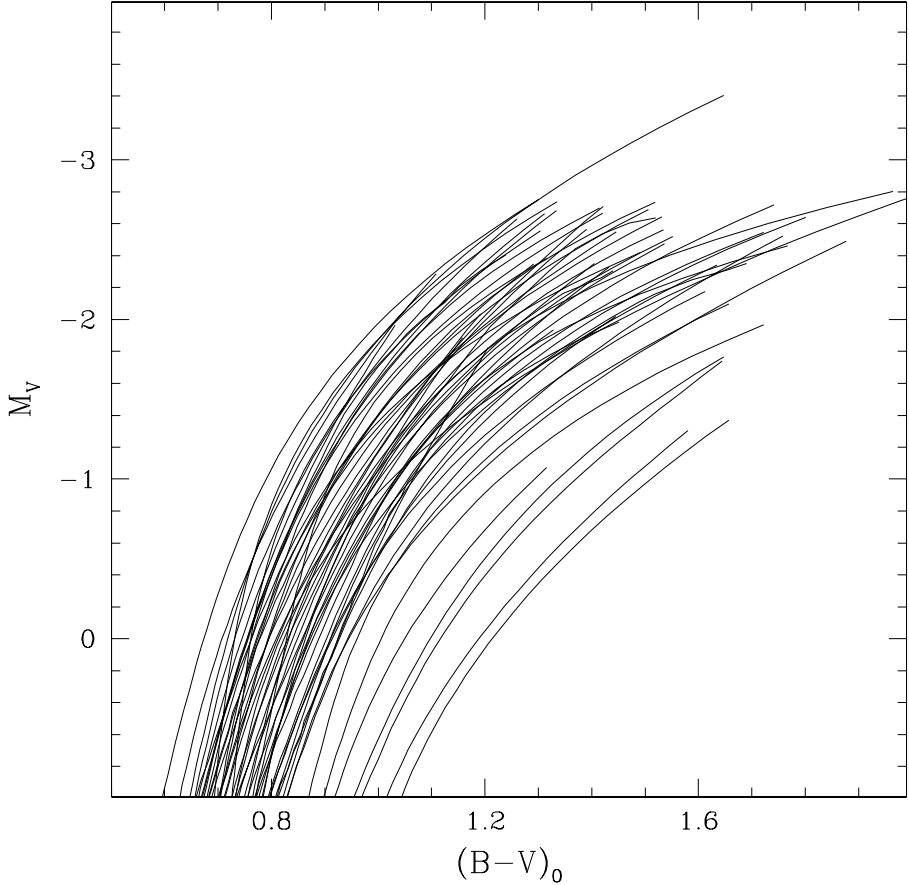}
\vspace{-3cm}
\caption{
\protect\label{lm}
RGB mean ridge lines for 55 GGCs in the absolute plane ($M_V,~(B-V)_0$).
The adopted DM as been computed assuming the metallicity in the CG97 scale 
(see text).
}
\end{figure*}

\begin{figure}
\epsffile{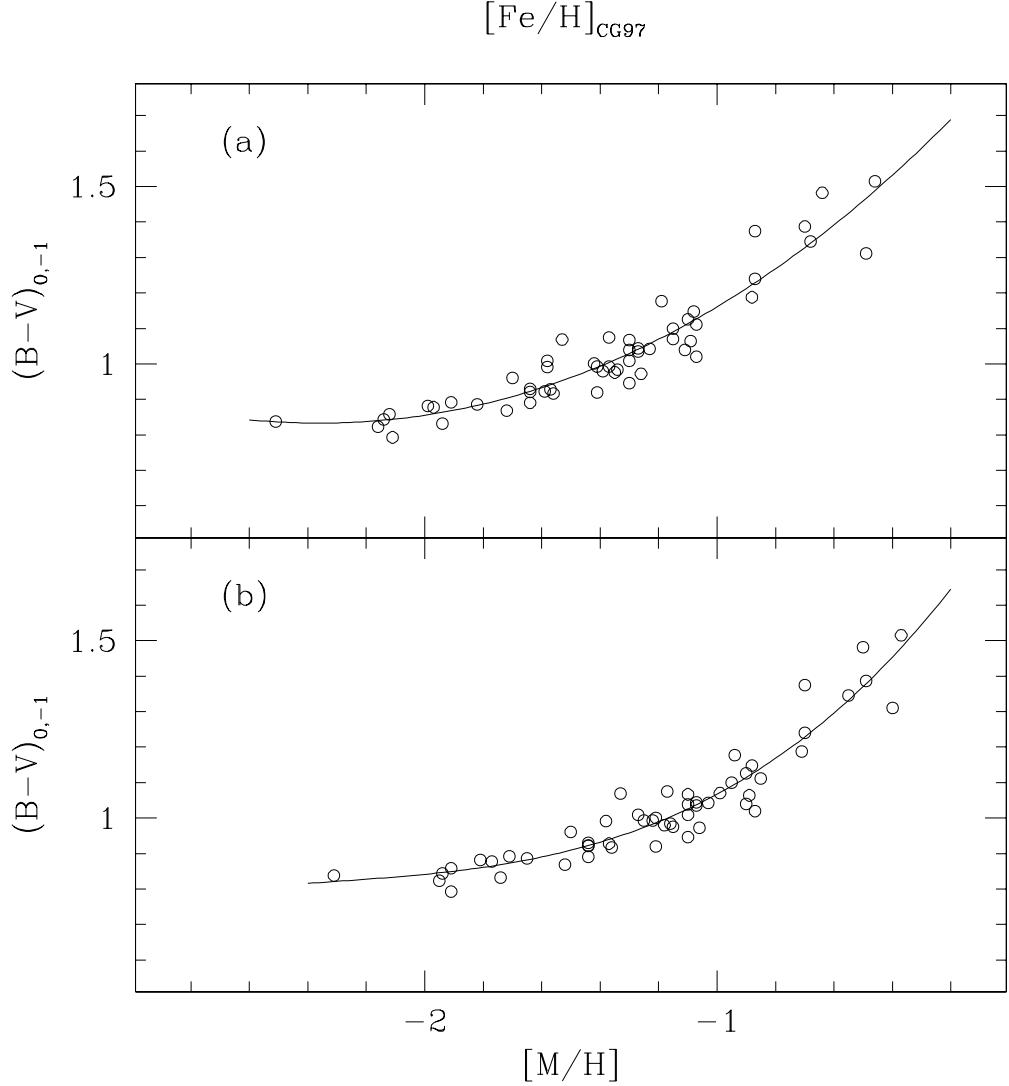}
\caption{
\protect\label{bv}
Intrinsic $(B-V)_0$ colour of the RGB, measured at $M_V=-1$
as function of $\feh_{\rm CG97}$ and $\mh$, {\it panel (a)} and {\it (b)},
respectively. The solid lines are the best fit relations listed in Table 4
(relation 4.15 and 4.16, respectively).
}
\end{figure}

\begin{figure}
\epsfxsize=5.0cm
\epsffile{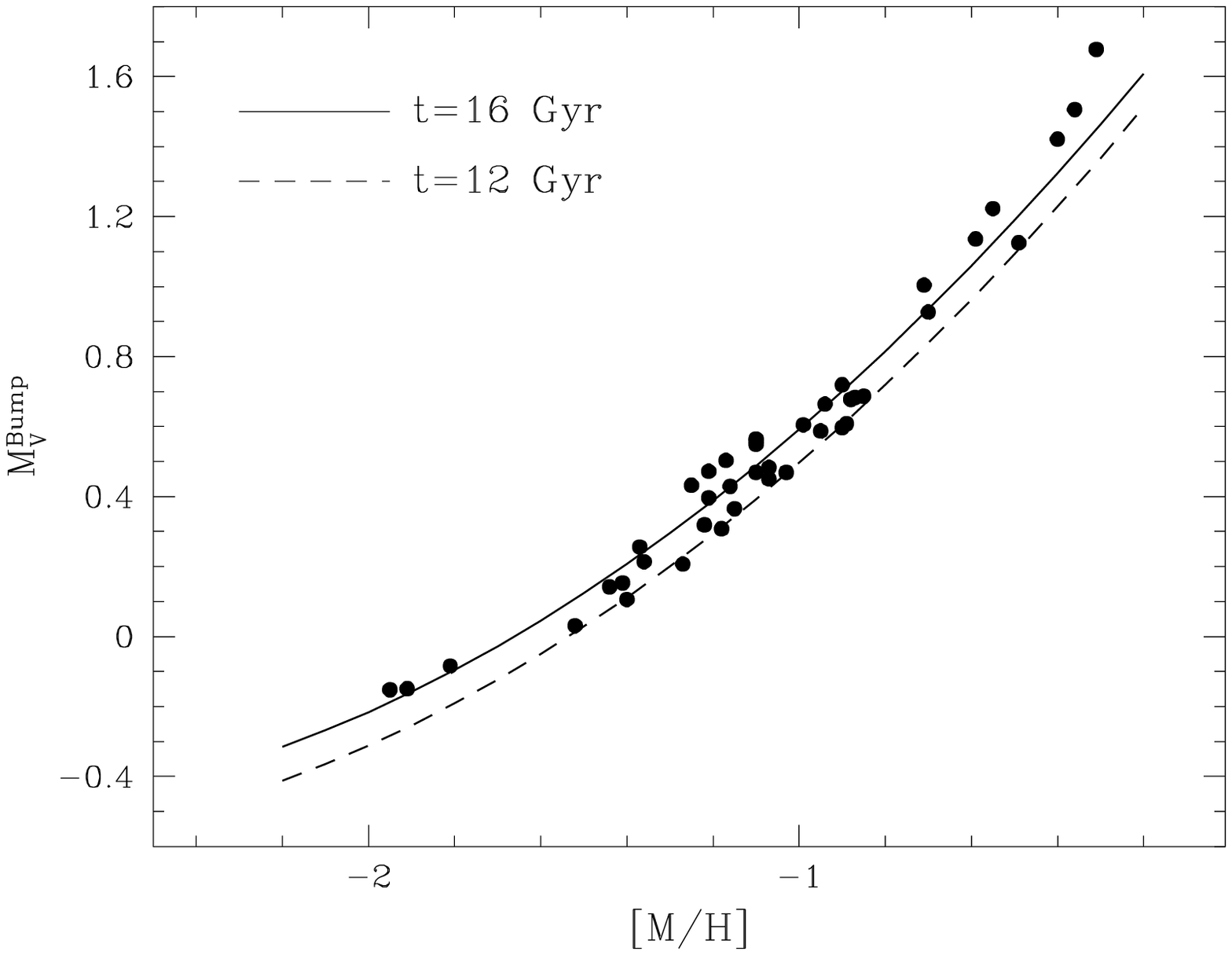}
\vspace{-5cm}
\caption{
\protect\label{bump5}
The absolute magnitude of the RGB Bump 
as a function of
the global metallicity $\mh$
(the DM has been computed accordingly).
The solid line
is the theoretical prediction
by CLS98 models at t=16 Gyr, the dashed line
represent the same set of models at t=12 Gyr. 
}
\end{figure}

 .

In Figure 16, we compare the theoretical and observed values of the
$\Delta V_{\rm HB}^{\rm AGB}$ parameter. The shaded region represents
the quoted uncertainty ($\pm 0.1$) in the absolute location of the AGB
clump.  Despite the quite large error bar affecting most of the (few)
available measurements of the AGB-clump, the level of the agreement
with the theoretical prediction is remarkable.  Such a result, especially
combined with that obtained in Section 6.4 for the RGB-bump location,
is comforting about the reliability and the internal consistency of
the adopted theoretical prescriptions.

For sake of completeness we give below the best fit relation of
the $\Delta V_{\rm HB}^{\rm AGB}$ parameter as a function of the
metallicity in the CG97 and {\it global} metallicity scale,
respectively:

\begin{equation}
\label{eq:4}
{\Delta V_{\rm HB}^{\rm AGB}} = { -0.16 \feh_{\rm CG97}-1.19
~~~~~~~~~~~~~~~~~(n=9,\sigma=0.06)}
\end{equation}

\begin{equation}
\label{eq:5}
{ \Delta V_{\rm HB}^{\rm AGB}} = { -0.17 \mh - 1.17
~~~~~~~~~~~~~~~~~(n=9,\sigma=0.05)}
\end{equation}

\section{TOWARDS ABSOLUTE QUANTITIES}

In order to carry out an exhaustive comparison with the models
we have to derive {\it absolute} quantities from our data. This
implies the knowledge (or the assumption) of a ``reliable'' distance
scale for the program clusters.  As it is well known (see for instance
the recent discussion by Gratton et al 1997, and Carretta et al 1999, 
hereafter C99)
different loci in the CMD and different standard candles can be
assumed to determine the distance to a given cluster (see also
Cacciari 1998 for an extensive review).
Here we adopt just {\it a given}
standard candle, and briefly comment on the possible impact of
alternative choices.  Any variation of the zero-point or of the
metallicity dependence of the luminosity of the adopted candles would
in fact affect the conclusions.

Since our study is mainly devoted to the quantitative analysis of the
evolved sequences (including the HB), it is quite natural to adopt HB
stars as standard candles, at least for heuristic
purposes. Unfortunately, although the HB is the {\it classical}
sequence traditionally used as reference branch, there is still strong
disagreement on the basic {\it absolute} calibration. For the sake of
discussion, in the following we will use our theoretical ZAHB as
standard candles, as this choice guarantees a complete
self-consistency in our approach. We leave to future studies the
assessment of the validity of the HB-models as suitable candles. We
discarded here the use of empirical relations since none of those
presented so far (see for references VandenBerg, Stetson \& Bolte,
1996, Gratton et al. 1997, Reid 1997, 1998, Cacciari 1998) 
actually calibrate the
``true'' ZAHB-level but rather adopt the mean apparent magnitude of
the HB at a given color, $<V_{\rm HB}>$, or an empirically derived
ZAHB level, obtained from the mean HB level via some metallicity and
HB morphology dependent correction factors.

The distance moduli were computed by adopting as reference equation (4) of
section 6.4.  Values were computed using both the metallicity scale
from the spectroscopic iron abundance measurements (GC97) and the
global metallicity scale (as derived in Section 3). Note that the
adoption of the CG97 spectroscopic scale rather than the Z85 leads to
an average decrease of the absolute luminosity of the ZAHB level and,
in turn, of the derived distance scale by $\sim 0.03$. An additional
decrease in luminosity ($\sim 0.04$ mag) occurs if
$\alpha-$enhancements are included to get the global metallicity
($\mh$).

The distance moduli obtained assuming $\feh_{\rm CG97}$ and $\mh$ are
listed in Table 2, column 7 and 8, respectively.  Note that in
computing the DM we adopted the individual reddening as listed in
column 5 of Table 2.  Considering that the derived DM are affected by
many uncertainties (namely, the evaluation of the ZAHB level, the zero
point and dependence on metallicity of the ZAHB level,
reddening, etc) we estimate that the global uncertainty affecting the
DM listed in Table 2 cannot be less than 0.2 mag.

These values can be then compared with those reported in the two most
recent compilations of GGC observable parameters: Djorgovski (1993)
and Harris (1996), respectively, which were, however, 
derived under
assumptions significantly different from those adopted here. In fact,
Djorgovski (1993) assumed a constant value for the HB level ($M_V^{\rm
HB} =0.6$) independent of metallicity, while Harris (1996) adopted $
M_V^{\rm HB} = 0.20 \feh + 1.0$, based on the empirical relation obtained
by Carney, Storm \& Jones (1992).

The residuals for the DM ($\rm this - previous~paper$) as a function of 
$\feh_{\rm CG97}$ are plotted in Figure 17a,b, for D93 and H96, respectively.
In the comparison with D93 there is a clear trend of the residuals
as a function of metallicity, mostly due to the assumption of a
constant $M_V^{\rm HB}$. No similar trend is detectable with respect to H96,
since  the assumption on the slope of the $M_V(HB)-\feh$ relation
is compatible with the one assumed here (see eq. 5). There is however a clear systematic 
offset ($\sim 0.15-0.2$) partially due to the different zero-point 
of the adopted relation and partially due to the difference in the 
procedure used to determine level of the ZAHB
(see Section 5).

\subsection{Comparison with other empirical distances}

As quoted in the previous section an extensive comparison between the
distance moduli obtained here and those derived adopting different
standard candles is beyond the purpose of the present paper. For sake
of example in this section we report two among the most recent results
obtained adopting different candles.

\subsubsection{The Main Sequence}

Among others, this approach was used by BCF89 who compared the
observed MS mean ridge lines for a sample of 19 GGCs with the {\it
reference} locus defined by 6 local subdwarfs. From this procedure
they derived distances and ages for the program clusters.  The
residuals of the comparison of the DM ($\rm this~paper-BCF89$) are plotted
as a function of metallicity ($\feh_{\rm CG97}$) in Figure 18a.  As
can be seen the mean difference is $\sim 0.15$ mag, the DM derived in
this paper being systematically larger than those obtained by BCF89.
The discrepancy ($\sim 0.6$) found for NGC 6809 is due in part to the
different photometry adopted here and in part to the different
assumption on the reddening: BCF89 assumed 0.14 while here we used
0.07 from Harris 1996).

The same methodological approach has been followed recently by Gratton
et al (1997), who gave new distances for a sample of 9 GGCs.  These
distances are based on high precision trigonometric parallaxes for a
sample of $\sim 30$ local subdwarf from the HIPPARCOS satellite.  They
found that the derived distances for the selected sample of GGCs are
systematically larger ($\sim 0.2$mag) than previously estimated.
The residuals of the corresponding distances ($\rm this~paper-
Hipparcos$) are plotted as a function of metallicity ($\feh_{\rm
CG97}$) in Figure 18b. 
While there is agreement between our DM and the
HIPPARCOS DM at the lowest metallicities, there seems to be a
systematically increasing discrepancy as metallicity increases.  The
sample is certainly too poor to derive any firm conclusion. 

However it is interesting to note that the recent
re-analysis of the HIPPARCOS data 
presented by C99
 goes in the direction of
showing a better agreement with the distances obtained
in this paper; in Figure 17c, the residuals with respect to
these most recent determinations have been plotted, as can be seen
 from the figure  the distances determinations for
clusters in the low-metallicity domain nicely agree,
while some systematic difference still remain at the high-metallicity end.
Since the difference on the distance modulus of about 0.07 mag
implies a corresponding difference in age of about $1 $ Gyr, it is quite 
evident that the differences found for specific cluster are still high
as far as the age determination is concerned. However the sample
is so small that it is still impossible to draw any reliable
conclusion.

It is also
important to bear in mind that due to the steepness of the MS the
``main sequence fitting'' method to derive distances is strongly
limited by any uncertainty affecting the MS colours (metallicity,
reddening, photometry calibration and the like) of both the clusters
and the reference stars (subdwarfs).

\subsubsection{White Dwarfs}

The cooling sequence of white dwarfs has been used recently by Renzini et 
al (1996) as a distance indicator to determine the distance of the nearby 
cluster NGC 6752. For this cluster they derived $(m-M)_0=13.05$ with an 
overall uncertainty of $\pm0.1$ mag. This value is compatible
with the DM obtained using the global metallicity ($\mh$)
$(m-M)_0=13.14\pm0.10$ (see column 8 in Table 2).

\subsection{The absolute quantities}

Using the assumptions and results obtained in the previous sections for 
the distances, it is possible to obtain various interesting plots
which describe in a very direct and clear way the properties of the RGB 
(and in particular its location and morphology) with varying metallicity.

In Figure 19, we present the mean ridge line for 55 GGCs in the absolute 
plane ($M_V,(B-V)_0$). The DM obtained from the $\feh_{\rm CG97}$ scale have
been adopted to construct the diagram.
 As can be noted by this figure at least three clusters (namely
NGC6333, NGC6535, NGC7492) appear to cross over the other
mean ridge lines suggesting that  photometries for these clusters
 could be affected by calibration problems and they deserve
a more accurate photometric analysis.

Figure 20a,b report the intrinsic colours of the RGB measured at $M_V=-1$
(labelled as $(B-V)_{0,-1}$ and listed in column 10
of Tbale 3) as a function of $\feh_{\rm CG97}$
and $\mh$, respectively.
 Also plotted are the 
best fit relations reported in Table 4 (eq. 4.15 and 4.16).

Figure 21 shows the dependence of the absolute location of the
RGB-Bump on the global metallicity. 
For
comparison with theoretical expectation, the relation by SCL97 has
been overplotted at two different ages (as in the previous figure), at
16 Gyr (the solid line) and 12 Gyr (the dashed line).  As can be seen
from this figure 
the previous discrepancy between the observation and
the model prediction for the location of this feature is completely
removed using the new models and considering the global metallicity
scale (as expected from the discussion in Section 6.4). 
Finally analytic
relations giving the absolute magnitude of the RGB-Bump as a function
of the metallicity using both the CG97 and {\it global} scale have
been computed. They are listed in Table 5 (relations 5.7 and
5.8). 

Though intriguing in principle, the uncertainty on the data is still
too large to allow any attempt to derive informations on a possible age spread
within the GGCs system from such a data-set.

\begin{figure*}
\epsfxsize=5.0cm
\epsffile{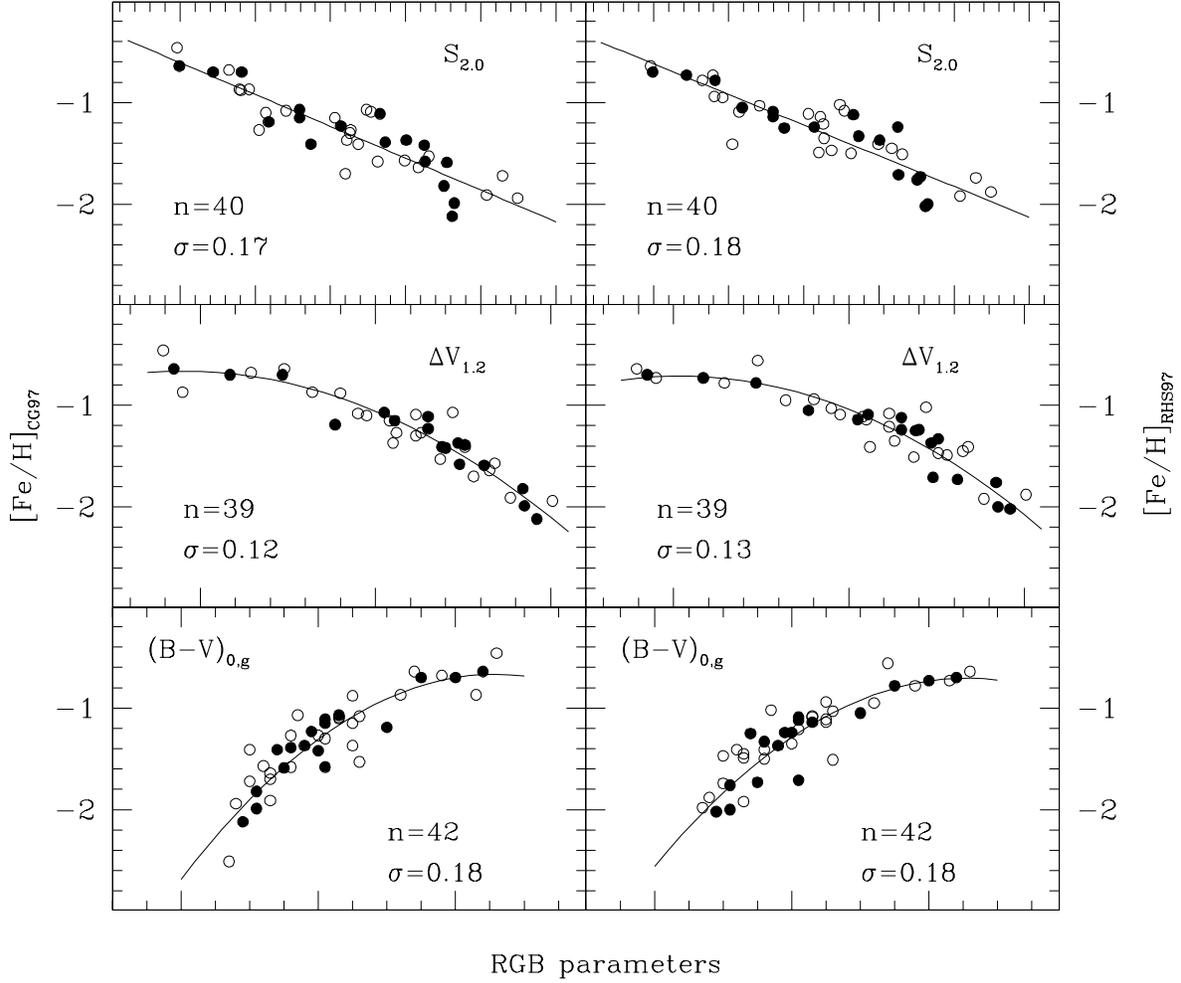}
\vspace{-2mm}
\caption{
\protect\label{}
$[Fe/H]_{CG97}$ versus the values obtained for three
RGB parameters defined in Section 6.2 
(namely $S_{2.0}$, $\Delta V_{1.2}$,
$(B-V)_{0,g}$) as computed in Section 3.2
({\it left panels}) and by RHS97 ({\it right panels}). 
The number of clusters used to compute each relation 
is reported, together with 
the standard deviations of the data.
In order to compute the scatter in an homogeneous way, only clusters
 in common with the RHS97 list have been considered in the 
left-panel plots.
}
\end{figure*}

\begin{figure*}
\epsfxsize=5.0cm
\epsffile{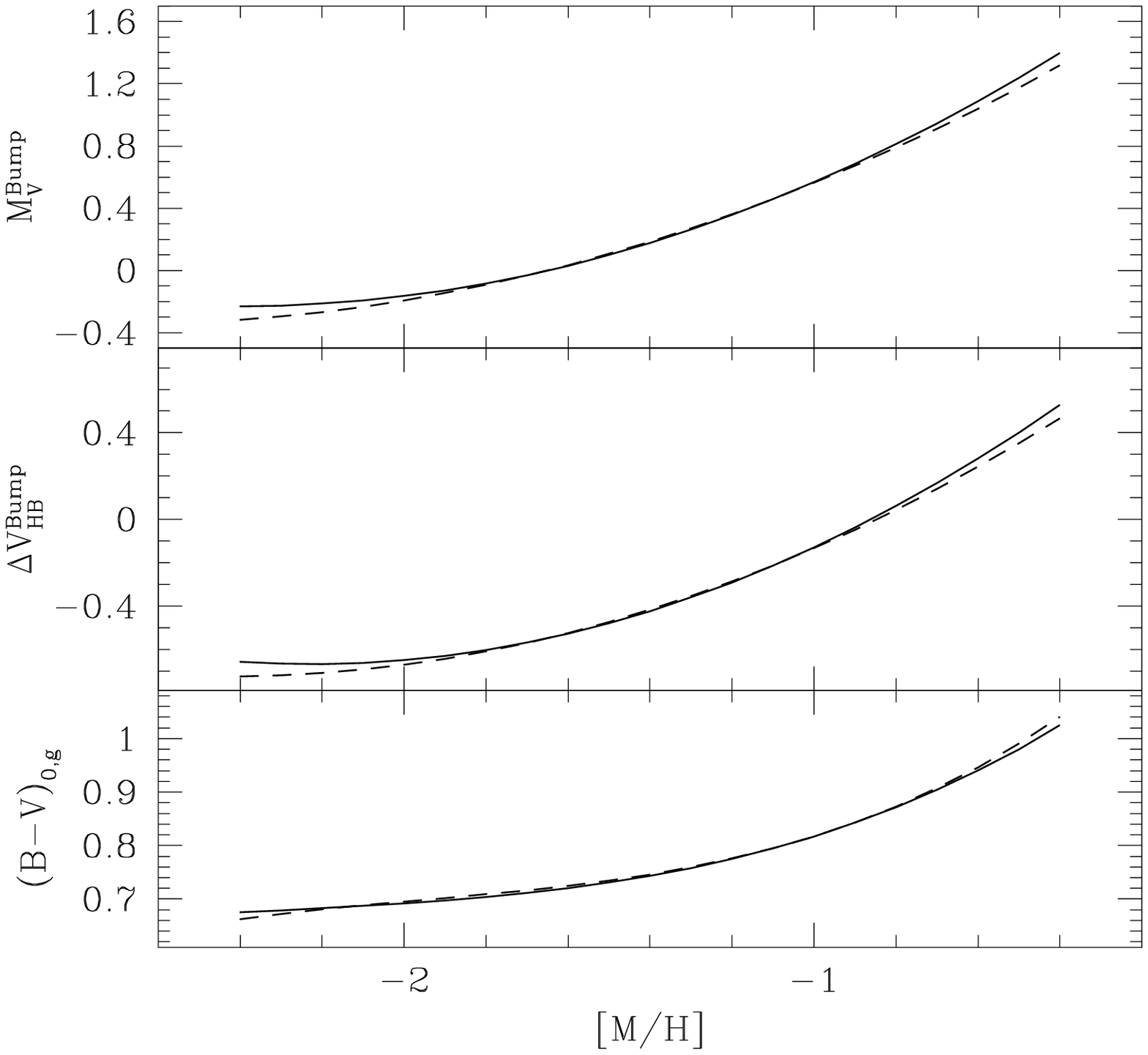}
\vspace{-1cm}
\caption{
\protect\label{} 
 $(B-V)_{0,g}$, $\Delta V_{HB}^{Bump}$
and $M_{V}^{Bump}$ as a function of the global metallicity,
computed adopting the $\alpha$ enhancement-relation 
plotted in Figure 1a 
({\it solid lines});
and adopting the $\alpha$ enhancement scenario proposed by Carney (1996)
 plotted in Figure 1b 
({\it dashed lines}).
As can be seen, the two different 
assumptions have a very small effect on the
derived relations. }
\end{figure*}

\section{FINAL REMARKS ON THE METALLICITY ASSUMPTIONS}

In this section we  briefly discuss the effects
of adopting different assumptions for the metallicity
 on the relations
we derived in the previous sections.
 In particular:

{\it (i)}
the use of the metal abundance ($[Fe/H]$) 
estimates obtained by RHS97 in the CG97
 scale instead of those listed in column 3 of Table 1;

{\it (ii)} the adoption of the Carney (1996) scenario for the
$\alpha-$element enhancing relation (see Figure 2b), rather than
that one plotted in Figure 2a, in computing the global metallicity.

{\it (i)} As already shown in Figure 1b the metal abundance 
obtained by RHS97 in the CG97 scale for the 42 GGCs in common
is fully compatible (within
 0.2 dex) with that one adopted in this paper
(computed following the procedure described in Section 3.2).
For this reason we expect very small effect on the relations 
we derived in the previous Sections. 
For sake of example in Figure 22 we report the results
for the relations we obtained for three of the RGB parameters
we defined in Section 6.2 (namely $S_{2.0}$, $\Delta V_{1.2}$,
$(B-V)_{0,g}$). On the left panels of Figure 21 we plotted
 the relations obtained assuming $[Fe/H]_{CG97}$ listed
 in Table 1, in the right panels those  assuming the values listed by RHS97.
The number of the clusters used to compute each relation is shown
in each panel together with the standard deviation of the data.
 In order to properly compare the scatter of the data with respect to
the best-fit relation
under the same assumptions,
only clusters in common between
 our sample and RHS97 have been used. 
As can be see from the comparisons between each couple
of panels the results are fully compatible
  both in terms of fit relations and data-scatter.

{\it (ii)} In panel (a) of Figure 2 we plotted the
$\alpha-$element enhancing relation  adopted to compute the
global metallicity listed in column 4 of Table 1.
However, as discussed in Section 3.4
 the trend of the 
$[\alpha/Fe]$ as a function of $[Fe/H]$, at least for GGCs,
 is still very uncertaint
expecially in the high-metallicity domain. For 
this reason   we  show the effects of 
 different assumptions in the $\alpha-$enhancing relation
on our results. 
We adopted the scenario proposed by Carney (1996)
 and plotted in Figure 2b. 
Even in this case for sake of example  in Figure 23 
we plotted three  relations obtained
in the previous Sections (respectively for $(B-V)_{0,g}$,
$\Delta V_{HB}^{Bump}$ and $M_V^{Bump}$) as a function
of the global metallicity computed adopting the two different scenarios.
In particular:
 {\it solid lines} are the best-fit relation obtained 
using global metallicity listed in Table 2,
while {\it dashed lines} are computed adopting
 the Carney (1996) scenario. As can be seen, only  a small effect
  is visible at the extreme ends
 of the relations, as  expected,
since the assumption of the Carney (1996) scenario only slightly increases
 (on average by 0.07 dex, the maximum beeing 0.13 dex) the global
 metallicity for metal rich (with $[Fe/H]_{CG97}<-1$)
 clusters (13 in our sample).

In summary, we can reasonably conclude that the relations derived in this paper
are little affected by the assumptions we adoped for the
metallicity.

\section{CONCLUSIONS AND FUTURE PROSPECTS}

A careful revision of all the best available CMDs
for the Post-Main Sequence branches (RGB,HB AGB) 
of GGCs has allowed to build a wide data-set.
A variety of observables quantitatively describing the main properties 
of the considered branches as far as the location and 
the basic features in the CMDs were measured:
the various quantities obtained via an homogeneous 
procedure applied to each individual CMD have been examined with varying 
the cluster metallicity, taking also into account the effects of
$\alpha-$enhancements and compared with the predictions of theoretical
models.
Very schematically this comparison has shown a substantial 
agreement between observations and theoretical
predictions, with a significant  improvement with respect to any 
similar previous study. The basic items which contribute to this result are: 
{\it (i)} the availability of a carefully tested wide sample of clusters (61);
{\it (ii)} the adoption of an innovative, homogeneous procedure to estimate 
the ZAHB level;
{\it (iii)} the adoption of new metallicity scales;
{\it (iv)} the comparison with up-dated, self-consistent models.

Further significant improvements in the analysis could be eventually done
as soon as new data for other clusters will be available and 
in particular when more accurate estimate on the global metal content
($\feh$ and $\alpha$-elements) of GGC stars
will be obtained via high resolution 
spectroscopy and new more accurate absolute distance moduli will be 
measured via alternative complementary methods.

\section{Acknowledgements}

We warmly thank a dear friend (Bob Rood) for a careful reading of the
manuscript.
The financial support of the ``Ministero della Universit\`a e della Ricerca
Scientifica e Tecnologica (MURST)'' to the project 
{\it Stellar Evolution} is kindly acknowledged.
FRF acknowledges the ESO visitor programm for its hospitality.

\end{document}